 \documentclass[twoside]{IEEEtran}
 \IEEEoverridecommandlockouts

\usepackage{array}
\usepackage{mdwmath}
\usepackage{mathtools}
\usepackage{subfig}
\usepackage{tabularx}
\usepackage{graphicx}
\usepackage{url}
\usepackage[noend]{algpseudocode}
\usepackage[ruled,vlined]{algorithm2e}
\graphicspath{{../pdf/}{../jpeg/}}
\DeclareGraphicsExtensions{.pdf,.jpeg,.png}

\usepackage{amsfonts}
\usepackage{booktabs}
\usepackage{lipsum}

\usepackage{tikz}
\usetikzlibrary{%
patterns,%
calc,%
fit,%
arrows,%
plotmarks,%
shadows,%
chains,%
shapes%
}
\usepackage{pgfplots}
\usepackage{cite}
\usepackage{amsmath,amssymb,amsfonts}
\newtheorem{lemma}{Lemma}
\newtheorem{theorem}{Theorem}
\newtheorem{remark}{Remark}
\newtheorem{corollary}{Corollary}

\usepackage{latexsym}
\usepackage{amsfonts,amssymb,amsmath}

\begin{document}

\title{Disruptive RIS for Enhancing Key Generation and Secret Transmission in Low-Entropy Environments}
\author{
\IEEEauthorblockN{Hibatallah Alwazani, \IEEEmembership{Student Member, IEEE}, Anas Chaaban, \IEEEmembership{Senior Member, IEEE}}\\

\thanks{%
The authors are with the School of Engineering, the University of British Columbia, 1137 Alumni Ave., Kelowna, BC V1V1V7, Canada (email: \{hibat97,anas.chaaban\}@ubc.ca)}

\thanks{{Part of this work was presented in \cite{alwazaniconf}.}}

}
\maketitle
\begin{abstract} 
 Key generation, a pillar in physical-layer security (PLS), {is the process of the exchanging signals from two legitimate users (Alice and Bob)} to extract a common key from the random, common channels. The drawback of extracting keys from wireless channels is the ample dependence on the dynamicity and fluctuations of the radio channel, rendering the key vulnerable to estimation by Eve (an illegitimate user) in low-entropy environments {because of insufficient randomness. Added to that, the lack of channel fluctuations} lower the secret key rate (SKR) defined as the number of bits of key generated per channel use. In this work, we aim to address this challenge by using a reconfigurable intelligent surface (RIS) to produce random phases at certain, carefully curated intervals such that it disrupts the channel in low-entropy environments.  
We propose an RIS assisted key generation protocol, study its performance, and compare with benchmarks to observe the benefit of using an RIS while considering various important metrics such as key mismatch rate and  secret key throughput. Furthermore, we characterize a scaling law { as a function of the rate of change of RIS phase switching} for the average secret information rate under this protocol. Then, we use both the key throughput and information rate to optimize the overall secrecy rate. Simulations are made to validate our theoretical findings and effectiveness of the proposed scheme showing an improvement in performance when an RIS is deployed.  
\end{abstract}


\section{Introduction}

\label{intro}
Physical layer security (PLS) provides means for information transmission that is provably secure. It is more desirable than classical cryptography \cite{cryptobook} for several reasons \cite{maurer}. First, there is no assumption made on the eavesdropper's computational power, they could have, in principle, unlimited computation power and still not be able to decipher the message in some scenarios. Second, it gives rise to the notion of perfect secrecy; knowledge of the ciphertext at the eavesdropper suggests nothing about the message being exchanged. {Finally, it is highly scalable \cite{plsRenzo}, which is necessary for future generation networks featuring devices with varying power and computation capabilities.}

The following question arises for key-based secret communication: How do two parties share a secret key without compromising the key? In PLS, a key can be extracted from the wireless channel which acts as a common and unique source of randomness for Alice (the transmitter) and Bob (the receiver). This idea of generating random, common, and secret keys at two legitimate points from wireless channels is not new \cite{keyreview,networkinfo}. It has been explored before but largely abandoned  because of the fundamental limitation of low-entropy environments where channels vary slowly thereby producing low secret key rates (SKR) \cite{unsuitableBits}. However, with the emerging concept of a smart radio environment enabled by the practical implementation of the reconfigurable intelligent surfaces (RIS)s, this limitation may be overcome \cite{smartRadio}. Here, the RIS is abstracted as a large number of passive, scattering elements, where each element can be reconfigured to change the amplitude and/or phase of the impinging electromagnetic (EM) waves to achieve a desired objective, such as inducing channel variations in our case. Essentially, the RIS becomes a basic building block for a programmable and software-defined wireless environment \cite{softwarecontrolIRS}. Thus, this {aspect of key generation from wireless channels} is undergoing a resurgence  in physical layer security (PLS) with a focus on smart radio environment enabled secret key generation {\cite{multiuser,mmWaveriskey,lowerboundSKR,chen2021intelligent,2024paper}}.

 When keys are generated from common random channels, the SKR depends on the entropy of the environment. For low mobility receivers which typically incur a low-entropy environment,  the coherence interval is relatively long and thus the SKR  might be small. It is better that the channel is dynamic, to increase the SKR. This can be accomplished by RISs via inducing randomness in the channel to resolve the caveat of static environments.


The authors in \cite{keygeneration} present a novel wireless key generation architecture based on randomized channel responses from an RIS which acts as the shared entropy source to Alice and Bob. {They conduct the first practical studies to  demonstrate an
RIS-based system achieving a key generation rate (KGR)
of 97.39 bps.}
The authors in \cite{multiuser} propose a joint user allocation scheme and an RIS reflection parameter adjustment scheme to enhance key generation efficiency in a multi-user communication scenario. They compare the result against a scheme with no RIS and find that the RIS indeed boosts performance by reducing channel similarities between adjacent users and thus enhancing the efficiency of key generation.

Furthermore, the authors in \cite{chen2021intelligent}  investigate key generation for an RIS assisted MISO channel, and the impact of channel correlation on the secret key rate. They acknowledge the effect of introducing an RIS into the system which can lead to key leakage since it brings about high correlation between legitimate users and eavesdroppers due to introducing a shared channel. They derive an upper bound on the SKR using {results} from \cite{maurer} and compare with another SKR upper bound without the presence of an RIS in the system {to showcase the gains one can obtain from utilising RISs}.
A study on the minimum achievable SKR in the presence of an RIS and multiple passive eavesdroppers is given in \cite{lowerboundSKR}, where the authors optimize the minimum SKR by choosing appropriate RIS phase shifts. Here, the RIS works in a capacity to combat deleterious wireless channel conditions such as co-channel interference and dead zones. Moreover, \cite{lowerboundSKR} considers an RIS as a new degree of freedom in the channel, where the aim of the RIS is increasing correlation between legitimate nodes' channels and decreasing correlation with eavesdropping channels.

While most existing works focus on enhancing channel randomness for key generation, a few studies, such as  \cite{otp}, delve into broader considerations encompassing secure data transmission in RIS-enabled setups.  The work in \cite{otp}  introduces a random phase shift scheme to induce environmental randomness and develops optimal solutions for time slot allocation.  

Earlier works consider idealistic assumptions such as perfect CSI  \cite{lowerboundSKR, otp}, do not account for spatial correlation in the RIS channels \cite{chen2021intelligent,multiuser,otp}, or do not examine {secrecy rate in favor of focusing only on the secret key rate}. Thus, it remains open to investigate practical implementations and  holistic solutions addressing those limitations, and to further explore the synergy between RIS and key generation mechanisms. In this paper, we address this gap by {proposing}  a generalized, practical scheme for key generation and optimizing {its performance in terms of secrecy rate}. 
{Our formulated optimization problem and solution encapsulates the work done in \cite{otp}. } Notably, our work diverges from aforementioned works by considering imperfect CSI, allowing for spatial correlation in the channel modelling, and {considering both key generation and secret transmission jointly in the analysis.} Furthermore, our work reveals how a {thorough channel estimation (CE) phase}
 may be unnecessary under a practical physical key generation scheme, and {hence can be avoided to reduce} time consumption and key leakage. {It also poses {and answers} an important question of: How much can an RIS help in boosting secrecy rate in this system?.
To address this,} we study secure communication in an RIS-assisted SISO system, leading to the following contributions:
\begin{itemize}
    \item {We propose a practical key generation scheme utilizing a spatially correlated RIS with switching random phase-shifts and highlight the effect of RIS parameters such as the number of elements and switching rate on performance.}
    \item We develop  a theoretical achievable SKR lower bound for the proposed scheme for correlated and uncorrelated Rayleigh fading channels.
    \item For a practical implementation of the protocol, we study the key mismatch rate (KMR) and the key throughput defined as the average number of key bits generated per transmission.
        \item {We derive the secrecy leakage incurred by Eve and derive the probability of successful attack to be accounted for in the final expression for secret transmission throughput. }
    \item We derive the scaling law for the average information rate and subsequently use it for the total secret transmission throughput for optimization.
    
\item { We formulate an optimization problem to increase the secret transmission throughput by finding optimal values for switching rate, key extraction  period, and channel estimates quantization. Furthermore, we scrutinize the significance of RIS size and location on system performance.}
\item {As a result of this work, we conclude that the secrecy rate is a function of the RIS and its characteristics (switching rate, size, inter-elements correlation, location, and phases) and subsequently provide a cohesive framework to analyse this dependence in detail.}

\end{itemize}

 Next, we formulate the system model studied in this paper.

\begin{figure}
    \centering
    \tikzset{every picture/.style={scale=1}, every node/.style={scale=1}}
    \begin{tikzpicture}
    \node (a) at (0,0) {};
    \node at ($(a)+(0,-.1)$) {\footnotesize Alice};
    \draw[line width=1] ($(a)+(-.2,.1666)$) to ($(a)+(0,.75)$) to ($(a)+(.2,.1666)$);
    \draw[line width=1] ($(a)+(.2,.1666)$) to ($(a)+(-.12,.3666)$);
    \draw[line width=1] ($(a)+(-.2,.1666)$) to ($(a)+(.12,.3666)$);

    \draw[line width=1] ($(a)+(.08,.5666)$) to ($(a)+(-.12,.3666)$);
    \draw[line width=1] ($(a)+(-.082,.5666)$) to ($(a)+(.12,.3666)$);
    \draw[line width=1.5] ($(a)+(-.09,.75)$) to  ($(a)+(.09,.75)$);
    \draw[line width=1.5] ($(a)+(0,.75)$) to  ($(a)+(0,.8)$);
    
	\node (b) at (6,-.1) {\ \ \footnotesize Bob};
	\draw[fill=gray!50!white] ($(b)+(0,.3)$) to ($(b)+(.25,.3)$) to ($(b)+(.3,.8)$) to ($(b)+(.05,.8)$) to ($(b)+(0,.3)$); 
	
	\node (e) at (3,-1.6) {\ \ \footnotesize Eve};
	\draw[fill=gray!50!red] ($(e)+(0,.3)$) to ($(e)+(.25,.3)$) to ($(e)+(.3,.8)$) to ($(e)+(.05,.8)$) to ($(e)+(0,.3)$); 
	    
	\node (r) at (3,3) {\footnotesize Rose};
	\node (r_corner) at ($(r)-(.75,.2)$) {};
    \draw[blue!50!white] (r_corner.center) to ($(r_corner.center)+(1.5,0)$) to 
    ($(r_corner.center)+(1.5,-1.5)$) to ($(r_corner.center)-(0,1.5)$) to (r_corner.center);
    \foreach \i in {1,..., 9}
    {
    \draw[blue!50!white] ($(r_corner.center)+(.15*\i,0)$) to ($(r_corner.center)+(.15*\i,-1.5)$);
    \draw[blue!50!white] ($(r_corner.center)+(0,-.15*\i)$) to ($(r_corner.center)+(1.5,-.15*\i)$);
    }
    
    \draw[->] ($(a)+(.2,.7)$) to node [above,sloped] {\footnotesize $\boldsymbol{h}_{\rm ar}$} ($(r)-(0,.95)$);
    \draw[->,dashed] ($(r)-(0,1.1)$) to node [below,sloped] {\footnotesize $\boldsymbol{h}_{\rm ra}$} ($(a)+(.2,.55)$);
    
    \draw[->,dashed] ($(b)+(-.2,.55)$) to node [below,sloped] {\footnotesize $\boldsymbol{h}_{\rm br}$} ($(r)-(-0.1,1.1)$);
    \draw[->] ($(r)-(-0.1,.95)$) to node [above,sloped] {\footnotesize $\boldsymbol{h}_{\rm rb}$} ($(b)+(-.2,.7)$);

    \draw[<-] ($(b)+(-.3,.45)$) to node [above,sloped] {\footnotesize $h_{\rm ab} \qquad \qquad $} ($(a)+(0.3,.45)$);
    \draw[<-,dashed] ($(a)+(0.3,.3)$) to node [below,sloped] {\footnotesize $h_{\rm ba} \qquad \qquad $} ($(b)+(-.3,.3)$);

    \draw[->] ($(a)+(0.3,.1)$) to node [below,sloped] {\footnotesize $h_{\rm ae}$} ($(e)+(-.2,.5)$);
    \draw[->,dashed] ($(b)+(-0.3,.1)$) to node [below,sloped] {\footnotesize $h_{\rm be}$} ($(e)+(.4,.5)$);
    \draw[->] ($(r)-(-0.05,1.2)$) to node [above,sloped] {\footnotesize $h_{\rm re}\qquad$} ($(e)+(.1,.9)$);

    \end{tikzpicture}
    \caption{System model with an access point Alice, receiver Bob, RIS Rose, and eavesdropper Eve.}
    \label{fig:sysmodel}
\end{figure}
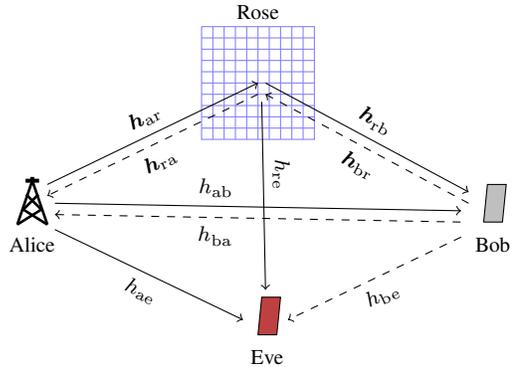

\section{System Model}
Consider the setup depicted in Fig. \ref{fig:sysmodel}, where a single antenna access point (Alice) serves a single-antenna user (Bob), in the presence of a passive eavesdropper (Eve) and an RIS (Rose) equipped with $N$ elements controlled by Alice through a control link. To secure the communication, Alice and Bob generate a key through exchanging signals over the wireless channel, and use the key to encrypt the message to be transmitted. Eve knows the crypto-system and the key generation protocol, and aims to discover the information exchanged between Alice and Bob.


As can be seen in Fig. \ref{fig:sysmodel}, we denote by ${h}_{\rm ab},h_{\rm ba}\in\mathbb{C}$ the channels from Alice to Bob and from Bob to Alice, respectively, by $h_{\rm ae},h_{\rm be}\in\mathbb{C}$ the channels from Alice and Bob to Eve, respectively, by $\boldsymbol{h}_{{\rm ar}},\boldsymbol{h}_{{\rm br}}  \in \mathbb{C}^{N}$ the channels from Alice and Bob to Rose, respectively, and by $\boldsymbol{h}_{\rm ra},\boldsymbol{h}_{\rm rb},\boldsymbol{h}_{\rm re} \in \mathbb{C}^{N}$ the channels from Rose to Alice, Bob, and Eve, respectively. 

We assume block-fading channels in which all channels maintain constant values for blocks of $T$ symbols and vary independently between blocks based on their respective distributions. We also assume a time-division duplexing (TDD) scheme, which implies that $h_{\rm ba}=h_{\rm ab}$, $\boldsymbol{h}_{\rm ra}=(\boldsymbol{h}_{\rm ar}^H)^T$, and $\boldsymbol{h}_{\rm rb}=(\boldsymbol{h}_{\rm br}^H)^T$ (reciprocal channels). Moreover, we assume correlated Rayleigh fading so that 
\begin{align}
\label{model_1}
&{h}_{ij}\sim \mathcal{CN}(0, \beta_{ij})\\
&\boldsymbol{h}_{{\rm r}i}\sim \mathcal{CN}(0, \beta_{{\rm r}i}\mathbf{R})
\end{align}
for $i,j \in \{{\rm a,b,e}\}$, where {$\beta_{ij},\beta_{{\rm r}i}, i,j \in \{{\rm a,b,e}\}$  are the path loss factors }and $\mathcal{CN}(0, \mathbf{R})$ denotes a circularly symmetric complex Gaussian distribution with mean $0$ and covariance matrix $\mathbf{R}$. 
According to \cite{correlationMatrixRIS}, the exact spatial correlation model under isotropic scattering for the RIS can be defined as 
\begin{align} 
\label{corrRIS}
[\mathbf{R}]_{m,n}={\rm sinc}\left( \frac{2\pi}{\lambda}\|\boldsymbol{u}_n-\boldsymbol{u}_m\|_2\right)
\end{align}
where $\lambda$ is the wavelength of operation, $\boldsymbol{u}_n=[ x_n,  y_n, z_n]$ is the location of element $n$ with respect to the origin, and $\|\boldsymbol{u}_n-\boldsymbol{u}_m\|_2$ is the distance between the $n^{th}$ and $m^{th}$ element. Note that for $\mathbf{R}=\mathbf{I}_{N}$ we have the special case for independent and identically distributed (i.i.d.) Rayleigh fading.

The transmission time in a block ($T$ symbols) is divided between Alice and Bob for the sake of key generation ($T_{\rm k}$ symbols) and information transmission ($T-T_{\rm k}$ symbols). 

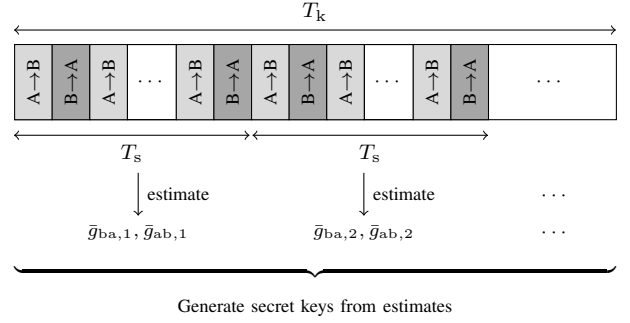
\begin{figure}
    \centering
\begin{tikzpicture}
\draw[<->] (0,1.2) to node[above]{\footnotesize $T_{\rm k}$} (8,1.2);
\node[anchor=south west,draw,minimum width = 8cm,minimum height = 1cm] at (0,0) {};

\node[anchor=south west,draw,minimum width = .5cm,minimum height = 1cm,fill=gray!30!white] at (0,0) {\rotatebox{90}{\scriptsize A$\to$B}};
\node[anchor=south west,draw,minimum width = .5cm,minimum height = 1cm,fill=gray!70!white] at (.5,0) {\rotatebox{90}{\scriptsize B$\to$A}};
\node[anchor=south west,draw,minimum width = .5cm,minimum height = 1cm,fill=gray!30!white] at (1,0) {\rotatebox{90}{\scriptsize A$\to$B}};
\node[anchor=south west,draw,minimum width = .65cm,minimum height = 1cm] at (1.5,0) {\scriptsize $\cdots$};
\node[anchor=south west,draw,minimum width = .5cm,minimum height = 1cm,fill=gray!30!white] at (2.15,0) {\rotatebox{90}{\scriptsize A$\to$B}};
\node[anchor=south west,draw,minimum width = .5cm,minimum height = 1cm,fill=gray!70!white] at (2.65,0) {\rotatebox{90}{\scriptsize B$\to$A}};
\draw[<->] (0,-.2) to node[below]{\footnotesize $T_{\rm s}$} (3.13,-.2);

\node[anchor=south west,draw,minimum width = .5cm,minimum height = 1cm,fill=gray!30!white] at (3.15,0) {\rotatebox{90}{\scriptsize A$\to$B}};
\node[anchor=south west,draw,minimum width = .5cm,minimum height = 1cm,fill=gray!70!white] at (3.65,0) {\rotatebox{90}{\scriptsize B$\to$A}};
\node[anchor=south west,draw,minimum width = .5cm,minimum height = 1cm,fill=gray!30!white] at (4.15,0) {\rotatebox{90}{\scriptsize A$\to$B}};
\node[anchor=south west,draw,minimum width = .65cm,minimum height = 1cm] at (4.65,0) {\scriptsize $\cdots$};
\node[anchor=south west,draw,minimum width = .5cm,minimum height = 1cm,fill=gray!30!white] at (5.3,0) {\rotatebox{90}{\scriptsize A$\to$B}};
\node[anchor=south west,draw,minimum width = .5cm,minimum height = 1cm,fill=gray!70!white] at (5.8,0) {\rotatebox{90}{\scriptsize B$\to$A}};
\draw[<->] (3.17,-.2) to node[below]{\footnotesize $T_{\rm s}$} (6.3,-.2);

\node[anchor=south west,minimum width = .65cm,minimum height = 1cm] at (6.8,0) {\scriptsize $\cdots$};

\node (gab1) at (1.65,-1.5) {\scriptsize $\bar{g}_{{\rm ba},1},\bar{g}_{{\rm ab},1}$};
\draw[->] (1.65,-.7) to node[right]{\scriptsize estimate} (gab1);

\node (gab2) at (4.65,-1.5) {\scriptsize $\bar{g}_{{\rm ba},2},\bar{g}_{{\rm ab},2}$};
\draw[->] (4.65,-.7) to node[right]{\scriptsize estimate} (gab2);
\node at (7.2,-1) {\scriptsize $\cdots$};
\node at (7.2,-1.5) {\scriptsize $\cdots$};

\node at (4,-2) {\scriptsize $\underbrace{\hspace{8cm}}$};
\node at (4,-2.5) {\scriptsize Generate secret keys from estimates};

\end{tikzpicture}
\caption{Key generation time $T_{\rm k}$ split into multiple RIS switching periods with duration $T_{\rm s}$ from each of which a channel estimate is obtained at Alice (A) denoted $\bar{g}_{{\rm ba}, \ell}$ and Bob (B) denoted $\bar{g}_{{\rm ab}, \ell}$. The estimates are then put through a process to generate a common key.}
    \label{fig:coherence}
\end{figure}

During the key generation phase, Alice transmits during odd time slots, while Bob transmits during even time slots as shown in Fig. \ref{fig:coherence}. Denoting the transmitted symbols by Alice and Bob by $x_{{\rm a},t}\in\mathbb{C}$ and $x_{{\rm b},t}\in\mathbb{C}$, respectively, which satisfy the power constraints $\sum_{t\text{ odd}} |x_{{\rm a},t}|^2\leq \frac{T_{\rm k}}{2}P$ and $\sum_{t\text{ even}} |x_{{\rm b},t}|^2\leq \frac{T_{\rm k}}{2}P$, the received signals can be written as
\begin{align}
{y}_{i,t} &=  (h_{{\rm a}i} + \boldsymbol{h}_{\rm ar}^H \boldsymbol{\Phi}_t \boldsymbol{h}_{{\rm r}i} ) x_{{\rm a},t} + z_{i,t}, \ \  i \in \{{\rm b, e}\},\ t \text{ odd,}\label{RxSignalBobandEve}\\
{y}_{i,t} &=  (h_{{\rm b}i} + \boldsymbol{h}_{\rm br}^H \boldsymbol{\Phi}_t \boldsymbol{h}_{{\rm r}i} ) x_{{\rm b},t} + z_{i,t}, \ \ i \in \{{\rm a, e}\},\ t \text{ even,}
\end{align}
where  
\begin{align}
\label{Phi_def}
\boldsymbol{\Phi}_t=\text{diag}([e^{j \theta_{t,1}}, \dots,e^{j\theta_{t,N}}])\in \mathbb{C}^{N\times N}
\end{align}
is the reflection matrix for Rose in time slot $t$, $\theta_{t,n}\in [0,2\pi]$ is the phase-shift applied by element $n$, and $z_{{\rm a},t},z_{{\rm b},t},z_{{\rm e},t}\in\mathbb{C}$ are noise samples at Alice, Bob, and Eve, respectively, which are independent of each other, and are i.i.d. over time with distribution $\mathcal{CN}(0,\sigma^2)$.
During this phase, Alice and Bob {aim to} generate a shared key $\boldsymbol{k}=(k_1,\ldots,k_r)\in\{0,1\}^r$ where $r$ is the number of key bits. The secret key rate (SKR) is defined as $R_{\rm k}=\frac{r}{T_{\rm k}/2}$ (in bits per transmission), which is desired to be large.

The end goal of Alice and Bob is to perform this key generation and extract $\boldsymbol{k}$ while preventing Eve from being able to discover the key.

{During the information transmission phase, we use $\boldsymbol{k}$ to encrypt Alice's message $\boldsymbol{m}$ to Bob in a certain way to ensure perfect secrecy.}

\section{Key Generation and Secret Key Rate}
Alice and Bob use the random channel between them as a source of shared randomness to generate a key. Since the channel remains constant during a coherence interval of length $T$ symbols, the RIS can help disrupt the channel by embedding additional randomness during a coherence interval. This is {\rm det}ailed next.
\subsection{Channel Estimation}
\label{subCE}
Alice and Bob generate keys by estimating their channels (Fig. \ref{fig:coherence}) and using the channel estimates as common randomness. To randomize the channel during a coherence interval, the RIS switches its phase-shift matrix $\boldsymbol{\Phi}_{\ell}$ randomly with  $\theta_{\ell,n}\sim\text{Uniform}[0,2\pi]$ every $T_{\rm s}$ symbols, with $T_{\rm s}\leq T_{\rm k}$ and $T_{\rm k}\mod  T_{\rm s} =0$. { Here, $\ell \in\{1,...,L\}$ where $L=T_{\rm k}/T_s$}. During switching interval $\ell$, Alice and Bob send pilot signals $\boldsymbol{x}_{{\rm a},\ell},\boldsymbol{x}_{{\rm b},\ell}\in\mathbb{C}^{T_{\rm s}/2}$ in odd-indexed and even-indexed symbols, respectively, such that $\|\boldsymbol{x}_{{\rm a},\ell}\|^2=\|\boldsymbol{x}_{{\rm a},\ell}\|^2=\frac{T_{\rm s}}{2}P$. Alice and Bob receive
\begin{align}
\boldsymbol{y}_{{\rm a},\ell}&=g_{{\rm ba},\ell}\boldsymbol{x}_{{\rm b},\ell}+\mathbf{z}_{{\rm a},\ell},\\
\boldsymbol{y}_{{\rm b},\ell}&=g_{{\rm ab},\ell}\boldsymbol{x}_{{\rm a},\ell}+\mathbf{z}_{{\rm b},\ell},
\end{align}
where $g_{{\rm ba},\ell}=h_{\rm ba} + \boldsymbol{h}_{\rm br}^H \boldsymbol{\Phi}_{\ell} \boldsymbol{h}_{\rm ra}$ and $g_{{\rm ab},\ell}=h_{\rm ab} + \boldsymbol{h}_{\rm ar}^H \boldsymbol{\Phi}_{\ell} \boldsymbol{h}_{\rm rb}=g_{{\rm ba},\ell}$ due to reciprocity, and $\mathbf{z}_{{\rm a},\ell}$ and $\mathbf{z}_{{\rm b},\ell}$ collect the noise instances during switching period $\ell$ during odd-indexed and even-indexed symbols, respectively. Alice and Bob then estimate the channels $g_{{\rm ba},\ell}$ and $g_{{\rm ab},\ell}$ to be used as shared randomness as follows (using least-squares estimation)
\begin{align}
\bar{g}_{{\rm ba},\ell}&=\boldsymbol{y}_{{\rm a},\ell}^H\frac{\boldsymbol{x}_{{\rm b},\ell}}{\|\boldsymbol{x}_{{\rm b},\ell}\|^2} = g_{{\rm ba},\ell} + \bar{z}_{{\rm ba},\ell},\label{Estimate_ba}\\
\bar{g}_{{\rm ab},\ell}&=\boldsymbol{y}_{{\rm b},\ell}^H\frac{\boldsymbol{x}_{{\rm a},\ell}}{\|\boldsymbol{x}_{{\rm a},\ell}\|^2} = g_{{\rm ab},\ell} + \bar{z}_{{\rm ab},\ell},\label{Estimate_ab}
\end{align}
where $\bar{z}_{{\rm ba},\ell}=\frac{\mathbf{z}_{{\rm a},\ell}^H\boldsymbol{x}_{{\rm b},\ell}}{\|\boldsymbol{x}_{{\rm b},\ell}\|^2}$ and $\bar{z}_{{\rm ab},\ell}=\frac{\mathbf{z}_{{\rm b},\ell}^H\boldsymbol{x}_{{\rm a},\ell}}{\|\boldsymbol{x}_{{\rm a},\ell}\|^2}$ are independent $\mathcal{CN}(0,\bar{\sigma}^2)$ noises with $\bar{\sigma}^2=\frac{2\sigma^2}{T_{\rm s}P}$. During the same time, Eve obtains the following estimates similarly
\begin{align}
\bar{g}_{{\rm be},\ell}&= g_{{\rm be},\ell} + \bar{z}_{{\rm be},\ell},\label{Estimate_be}\\
\bar{g}_{{\rm ae},\ell}&= g_{{\rm ae},\ell} + \bar{z}_{{\rm ae},\ell},
\end{align}
where $g_{{\rm be},\ell}=h_{\rm be} + \boldsymbol{h}_{\rm br}^H \boldsymbol{\Phi}_{\ell} \boldsymbol{h}_{\rm re}$, $g_{{\rm ae},\ell}=h_{\rm ae} + \boldsymbol{h}_{\rm ar}^H \boldsymbol{\Phi}_{\ell} \boldsymbol{h}_{\rm re}$, and $\bar{z}_{{\rm be},\ell}$ and $\bar{z}_{{\rm ae},\ell}$ are independent $\mathcal{CN}(0,\bar{\sigma}^2)$ noises.

Next, Alice and Bob use the estimates $\bar{g}_{{\rm ba},\ell}$ and $\bar{g}_{{\rm ab},\ell}$ (which are dependent) to generate their shared key (as in \cite[Ch. 22]{elgamal_kim}). Note that since the channels are assumed to not vary during a coherence block and to vary between blocks, the channel estimates spanning multiple coherence blocks will  be independent but  not identically distributed due to {$\boldsymbol{\Phi}_{\ell} $}. However, Alice and Bob can use the $\ell^{\rm th}$ switching period in \textit{multiple coherence blocks} to generate one key, making the estimates used to generate a given key i.i.d. as desired. This means that each coherent block contributes to bits in $\frac{T_{\rm k}}{T_{\rm s}}$ keys that are generated by collecting bits from multiple coherence blocks.

Next, we introduce channel correlation into the system.
\subsection{Channel Correlation}
The Pearson correlation coefficient $\rho$ between two fading channels from a single transmitter to two receivers separated by a distance $d$ is calculated via
 \begin{align}
     \label{pearson}
     \rho=\left[J_0\big(\frac{2\pi d}{\lambda}\big)\right]^2,
 \end{align}
 where $J_0(\cdot)$ is the Bessel function of the first kind.

Specifically,  $\rho$ in \eqref{pearson}  is the correlation between the direct channels $h_{\rm ab},h_{ae}$ (Alice to Bob and Alice to Eve). Also, \eqref{pearson} describes correlations between components of the vector channels from the RIS to Bob and to Eve, $\boldsymbol{h}_{\rm rb},\boldsymbol{h}_{\rm re}$, respectively.

{Usually, this correlation is often ignored in the literature \cite{pearson} and the amount of leaked information to Eve is considered negligible}. For instance,  if the distance between Bob and Eve is greater than or equal to half of the wavelength $\frac{\lambda}{2}$, for example $5$ cm {at the carrier frequency of $3$ GHz}, they will experience nearly uncorrelated fading channels from Alice. This means that the fading channels observed by Bob and Eve will be uncorrelated, making it difficult for Eve to extract information about the secret key generated by Bob.    {Despite this}, this works considers that Eve is closer to Bob than Alice and accounts for the spatial correlation between the channels of Alice to Bob and Alice to Eve and the resulting secrecy leakage { for a more comprehensive analysis.} The special case where the channels are uncorrelated  is explored later on in the mathematical derivations and simulations as a meaningful benchmark.
In the next section, we define the SKR.

\subsection{Secret Key Rate}
Let the estimates $\bar{g}_{{\rm ba},\ell}$, $\bar{g}_{{\rm ab},\ell}$, $\bar{g}_{{\rm be},\ell}$ and $\bar{g}_{{\rm ae},\ell}$ be represented by random variables $\bar{G}_{{\rm ba}}$, $\bar{G}_{{\rm ab}}$, $\bar{G}_{{\rm be}}$ and $\bar{G}_{{\rm ae}}$, respectively. The SKR can be lower bounded by 
$$R_{\rm k}\geq  \frac{1}{T_{\rm s}/2} R_{\rm k}^{\rm lb},$$ where \cite{maurer} 
\begin{align}\label{SKR_lb}
R_{\rm k}^{\rm lb}&\triangleq I(\bar{G}_{\rm ab}; \bar{G}_{\rm ba})\\
&\qquad-\min\{I(\bar{G}_{\rm ab};\bar{G}_{\rm ae}),I(\bar{G}_{\rm ba};\bar{G}_{\rm ae})\}\nonumber\\
&=-{\sf h}(\bar{G}_{\rm ab}| \bar{G}_{\rm ba})\nonumber\\
&\qquad+\max\{{\sf h}(\bar{G}_{\rm ab}|\bar{G}_{\rm ae}),{\sf h}(\bar{G}_{\rm ba}|\bar{G}_{\rm ae})\},\nonumber
\end{align}
$I(X;Y)$ is the mutual information, ${\sf h}(X|Y)$ is the conditional entropy. 
 To simplify this lower bound, we need to study the distributions of the channel estimates. 
We first note that $\bar{G}_{{\rm ba}}$, $\bar{G}_{{\rm ab}}$, and $\bar{G}_{{\rm ae}}$ have zero mean. {Denote  ${\rho}_{\rm ab}=\beta_{\rm ab}+\beta_{\rm ar}\beta_{\rm rb}\|\mathbf{R}\|_F^2$ and ${\rho}_{\rm ae}=\beta_{\rm ae}+\beta_{\rm ar}\beta_{\rm re}\|\mathbf{R}\|_F^2$, then} the covariance of the channels are given by
\begin{align}
\tilde{\rho}_{\rm ab}&=\mathbb{E}[\bar{G}_{\rm ab}\bar{G}_{\rm ab}^*]=\mathbb{E}[\bar{G}_{\rm ab}\bar{G}_{\rm ba}^*]=\mathbb{E}[\bar{G}_{\rm ba}\bar{G}_{\rm ba}^*]\nonumber\\
&={\rho}_{\rm ab}+\bar{\sigma}^2,\\
\tilde{\rho}_{\rm ae}&=\mathbb{E}[\bar{G}_{\rm ae}\bar{G}_{\rm ae}^*]={\rho}_{\rm ae}+\bar{\sigma}^2,\\
\tilde{\rho}_{\rm ab}^{\rm ae}&=\mathbb{E}[\bar{G}_{\rm ab}\bar{G}_{\rm ae}^*]=\mathbb{E}[\bar{G}_{\rm ba}\bar{G}_{\rm ae}^*]\\&=\rho(\sqrt{\beta_{\rm ab}\beta_{\rm ae}}+\beta_{\rm ar}\sqrt{\beta_{\rm rb}\beta_{\rm re}}\|\mathbf{R}\|_F^2).
\end{align}
Then, 
\begin{align}
\boldsymbol{\Sigma}_{\rm ab}={\rm Cov}\left([\bar{G}_{\rm ab}\ \bar{G}_{\rm ba}]^T\right)
&=\begin{bmatrix} \rho_{\rm ab}+\bar{\sigma}^2 & \rho_{\rm ab}\\
\rho_{\rm ab} & \rho_{\rm ab}+\bar{\sigma}^2\end{bmatrix},\label{Covab}\\
\boldsymbol{\Sigma}_{\rm abe}={\rm Cov}\left([\bar{G}_{\rm ab}\ \bar{G}_{\rm ae} ]^T\right)&={\rm Cov}\left([\bar{G}_{\rm ba}\ \bar{G}_{\rm ae} ]^T\right)\label{Covabe}\\
&\hspace{-1cm}=\begin{bmatrix}
\tilde{\rho}_{\rm ab} & \tilde{\rho}_{\rm ab}^{\rm ae} \\
\tilde{\rho}_{\rm ab}^{\rm ae} & \tilde{\rho}_{\rm ae}
\end{bmatrix}.\nonumber
\end{align}
Finally, we need the following statement to characterize the distribution of the estimates. 

\begin{lemma}\label{LemmaGaussian}
Given circularly symmetric complex Gaussian channels, the aggregate channel $g_{{ij},\ell}=h_{ij}+\boldsymbol{h}_{i{\rm r}}^H\boldsymbol{\Phi}_{\ell}\boldsymbol{h}_{{\rm r}j}$, $i\in\{{\rm a,b}\}$, $j\in\{{\rm a,b,e}\}$, $j\neq i$, can be modeled as a circularly symmetric complex Gaussian when $N$ is large.
\end{lemma}
\begin{IEEEproof}
The statement is a consequence of the central limit theorem, and can be proved similar to \cite[Lemma 2]{impactofPhaseRandomIRS}.
\end{IEEEproof}

Based on this, the lower bound $R_{\rm k}^{\rm lb}$ can be simplified as follows.

\begin{theorem} 
\label{theorrkmlb}
The SKR lower bound in \eqref{SKR_lb} simplifies to
\begin{align}
\label{rkmlb}
R_{\rm k}^{\rm lb}=\frac{1}{T_{\rm s}/2}\log_2\left(\frac{(\tilde{\rho}_{\rm ab}\tilde{\rho}_{\rm ae}-{\rho}_{\rm ab}^{{ae}^2})\tilde{\rho}_{\rm ab}}{\bar{\sigma}^2(2\rho_{\rm ab}+\bar{\sigma}^2)\tilde{\rho}_{\rm ae}}\right).
\end{align}
\end{theorem}
\begin{IEEEproof}
The statement is obtained by evaluating \eqref{SKR_lb} with circularly symmetric complex Gaussian channel estimates (using Lemma \ref{LemmaGaussian}) and obtaining
\begin{align}
\label{rklbcorr}
R_{\rm k}^{\rm lb}=\frac{1}{T_{\rm s}/2}\log_2\left(\frac{{\rm det}(\boldsymbol{\Sigma}_{abe})\tilde{\rho}_{\rm ab}}{{\rm det}(\boldsymbol{\Sigma}_{ab})\tilde{\rho}_{\rm ae}}\right),
\end{align}
then plugging the covariance matrices in \eqref{Covab} and \eqref{Covabe}.
\end{IEEEproof}

\begin{corollary}
\label{corrlskr}
The SKR lower bound in \eqref{SKR_lb} under uncorrelated {channels between Bob and Eve, i.e.,   $\rho=0$, }  becomes
\begin{align}
\label{rkmlbuncor}
R_{\rm k}^{\rm lb}=\frac{1}{T_{\rm s}/2}\log_2\left(1+\frac{\rho_{\rm ab}^2}{\bar{\sigma}^2(2\rho_{\rm ab}+\bar{\sigma}^2)}\right).
\end{align}
\end{corollary} 
\begin{IEEEproof}
Similar to Theorem \ref{theorrkmlb}, but 
with zero channel correlations, the covariance matrix in \eqref{Covabe} becomes
\begin{align}
\label{Covabe2}
\boldsymbol{\Sigma}_{abe}=\begin{bmatrix}
\tilde{\rho}_{\rm ab}   & 0\\
0 & \tilde{\rho}_{\rm ae} 
\end{bmatrix}.\nonumber\end{align}
Substituting in \eqref{rklbcorr} leads to \eqref{rkmlbuncor}.

\end{IEEEproof}

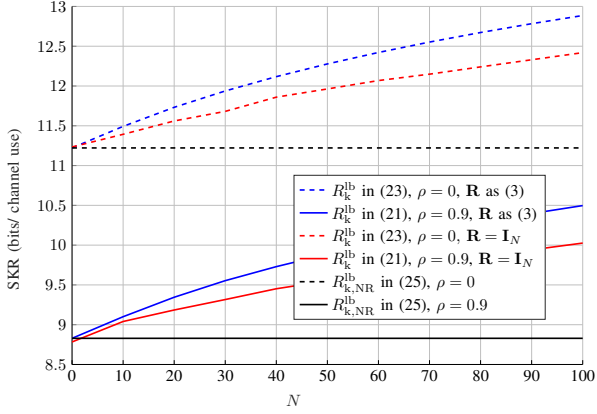
\begin{figure}[t]
\centering
\tikzset{every picture/.style={scale=0.62}, every node/.style={scale=1}}
%
%
\begin{tikzpicture}

\begin{axis}[%
width=6.953in,
height=4.879in,
at={(1.166in,0.658in)},
scale only axis,
xmin=0,
xmax=100,
xtick style={color=black},
xtick={0,10,20,30,40,50,60,70,80,90,100},
xticklabels={0,10,20,30,40,50,60,70,80,90,100},
xlabel style={at={(0.7,0)},font=\color{white!15!black}},
xlabel={$N$},
ymin=8.5,
ymax=13,
ylabel style={at={(0,0.7)},font=\color{white!15!black}},
ylabel={SKR (bits/ channel use)},
axis background/.style={fill=white},
axis x line*=bottom,
axis y line*=left,
xmajorgrids,
ymajorgrids,
legend style={at={(0.7,0.202)}, anchor=south west, legend cell align=left, align=left, draw=white!15!black}
]
\addplot [color=blue, dashed, line width=1.0pt]
  table[row sep=crcr]{%
0	11.2217103336092\\
10	11.4925026348278\\
20	11.7322394160411\\
30	11.9377572909673\\
40	12.117615656916\\
50	12.2775164903638\\
60	12.42144979743\\
70	12.5523170285508\\
80	12.6722942522178\\
90	12.783055464079\\
100	12.8859160711858\\
};
\addlegendentry{$R_{\rm k}^{\rm lb}$ in  \eqref{rkmlbuncor}, $\rho=0$, $\mathbf{R}$ as \eqref{corrRIS}}

\addplot [color=blue, line width=1.0pt]
  table[row sep=crcr]{%
0	8.82892333419396\\
10	9.10046674451751\\
20	9.34511168367865\\
30	9.55365884560813\\
40	9.72983800759579\\
50	9.88586362062249\\
60	10.0308713010523\\
70	10.1644304726055\\
80	10.2837230036859\\
90	10.3971503410498\\
100	10.4970913956306\\
};
\addlegendentry{ $R_{\rm k}^{\rm lb}$ in \eqref{rkmlb}, $\rho=0.9$, $\mathbf{R}$ as \eqref{corrRIS}}

\addplot [color=red,  dashed,line width=1.0pt]
  table[row sep=crcr]{%
0	11.2338050468875\\
10	11.393020301931\\
20	11.5606032939838\\
30	11.6812040396556\\
40	11.8597552862475\\
50	11.9630475860478\\
60	12.0684308058674\\
70	12.1488721978552\\
80	12.2403084117806\\
90	12.3297673730484\\
100	12.4192600012282\\
};
\addlegendentry{$R_{\rm k}^{\rm lb}$ in \eqref{rkmlbuncor}, $\rho=0,$ $\mathbf{R}=\mathbf{I}_N$ }

\addplot [color=red, line width=1.0pt]
  table[row sep=crcr]{%
0	8.78327660467636\\
10	9.04027083541813\\
20	9.1847373615036\\
30	9.31610246265417\\
40	9.45180488721633\\
50	9.55105929424256\\
60	9.71254141152402\\
70	9.73113227716363\\
80	9.82990138995566\\
90	9.94090982497089\\
100	10.0259698113884\\
};
\addlegendentry{$R_{\rm k}^{\rm lb}$ in  \eqref{rkmlb}, $\rho=0.9$, $\mathbf{R}=\mathbf{I}_N$}

\addplot [color=black, dashed, line width=1.0pt]
  table[row sep=crcr]{%
0	11.2217103336092\\
10	11.2217103336092\\
20	11.2217103336092\\
30	11.2217103336092\\
40	11.2217103336092\\
50	11.2217103336092\\
60	11.2217103336092\\
70	11.2217103336092\\
80	11.2217103336092\\
90	11.2217103336092\\
100	11.2217103336092\\
};
\addlegendentry{ $R_{\rm k,NR}^{\rm lb}$ in \eqref{rkmlbNoRIS}, $\rho=0$}

\addplot [color=black, line width=1.0pt]
  table[row sep=crcr]{%
0	8.82892333419396\\
10	8.82864574242352\\
20	8.82889429153985\\
30	8.82899427018555\\
40	8.82884416077455\\
50	8.82866347109046\\
60	8.82873290808089\\
70	8.82887441826368\\
80	8.82885824921212\\
90	8.82899595592446\\
100	8.82887696670495\\
};
\addlegendentry{ $R_{\rm k,NR}^{\rm lb}$ in \eqref{rkmlbNoRIS}, $\rho=0.9$}
\end{axis}

\begin{axis}[%
width=8.972in,
height=5.986in,
at={(0in,0in)},
scale only axis,
xmin=0,
xmax=1,
ymin=0,
ymax=1,
axis line style={draw=none},
ticks=none,
axis x line*=bottom,
axis y line*=left
]
\end{axis}
\end{tikzpicture}%
    \caption{{Theoretical SKR against $N$ with no RIS as a benchmark. The effects of  the correlation coefficient $\rho$ between Bob and Eve and the RIS  correlation matrix $\mathbf{R}$ are highlighted.}}
    \label{fig:LBskrvsN}
\end{figure}

   
\begin{corollary} 
Let 
 $h_{\rm ab}, h_{\rm ba}, h_{\rm ae}$ be represented by random variables  $H_{\rm ab}, H_{\rm ba}, H_{\rm ae}$.  Then, the SKR without an RIS is
\begin{align}
    R_{\rm k, NR}^{\rm lb}&={\sf h}(H_{\rm ab}|H_{ae})-{\sf h}(H_{\rm ab}|H_{ba}),\\
   \label{rkmlbNoRIS} &=
\frac{1}{T_{\rm s}/2}\log_2\left(\frac{(\tilde{\beta}_{\rm ab}\tilde{\beta}_{\rm ae}-\rho^2\beta_{\rm ab}\beta_{\rm ae})\tilde{\beta}_{\rm ab}}{(2\beta_{\rm ab}+\bar{\sigma}^2)\bar{\sigma}^2\tilde{\beta}_{\rm ae}}\right),
\end{align}
where $\tilde{\beta}_{\rm ab}=\beta_{\rm ab}+\bar{\sigma}^2$ and $\tilde{\beta}_{\rm ae}=\beta_{\rm ae}+\bar{\sigma}^2 $.
\end{corollary}
\begin{IEEEproof}
{Similar to Theorem \ref{theorrkmlb}, by setting $\beta_{\rm ar}$ to zero.}
\end{IEEEproof}

\begin{table}[!t]
\centering
\footnotesize
\caption{Simulation parameters.}
\scalebox{1}{\begin{tabular}{|l|l|}
\hline
  \textbf{Parameter} & \textbf{Value} \\ 
\hline
\textbf{Array parameters:} & \\
\hline
Carrier frequency & $3$ GHz \\
RIS configuration & Uniform planar array {$N=N_cN_r$}\\
Elements of RIS   & $N=100,N_c=20, N_r=5, \frac{\lambda}{4} $ spacing\\
 Antenna gain & 5dBi\\
 $d_{IRS}$& $0.5\lambda$\\
Noise power $\sigma^2$ & $-96\rm{dBm}$ \\
$P$ & $25\rm{dBm}$ \\
 \hline
\textbf{Path Loss:} & \\
\hline
Model & $\frac{10^{-C/10}}{d^{\alpha}}$ \\
$C$ (Fixed loss at $d=1$m)  & $30$\rm{dB} \\
$\alpha$ (Path loss exponent) & $2.2$ ($\beta_{\rm ar},\beta_{\rm re},\beta_{\rm rb}$), $3.67$ ($\beta_{\rm ab},\beta_{\rm ae},\beta_{\rm be}$)\\
\hline
\textbf{System Configuration:} & \\
\hline
$(F,T,N)$ & $(100,1000,100)$ \\
\hline
\end{tabular}}
\label{tabfigu}
\label{tabsims}
\end{table}

{
To visualize the effect of channel correlations 
be it from the proximity of Eve to Bob or from the closeness of the RIS elements on the SKR, we  evaluate the expressions for \eqref{rkmlb}, \eqref{rkmlbuncor}, and \eqref{rkmlbNoRIS} in 
Fig. \ref{fig:LBskrvsN}. We simulate the system with parameters provided in Table \ref{tabsims}. We consider Alice located at $(0,0,1.5)m$, RIS's center point deployed at {$(0,3,1.5)m$}, Bob located at $(30,0,1.5)m$, and Eve's position is {at ($x_{Eve},0,1.5$)m where $x_{Eve}=30+\frac{\lambda}{10}$ or $x_{Eve}=30+10\lambda$ corresponding to  $\rho=0.9$ or $\rho=0$, respectively. }Path loss factors are computed at {$3$} GHz for the 3GPP Urban Micro (UMi) scenario from TR36.814 (cf. \cite[Sec. V]{nadeem2020intelligent} ).}

{ The following several observations can be made: Assuming no spatial correlations at the RIS, i.e., $\mathbf{R}=\boldsymbol{I}_N$ lowers the overall SKR compared to considering a spatially correlated channel model for the RIS.  The reasoning behind this is as the RIS inter-element spacing decreases to less than $\frac{\lambda}{2}$, the entries of the correlation matrix  $\mathbf{R}$ increase and so does $\|\mathbf{R}\|_F^2$ {which increases the correlation coefficients}. This phenomenon is explored in more detail in \cite{spatiallycorr,2024paper}. One can conclude {that it is desirable to have temporally uncorrelated channels and spatially correlated RIS channels to increase SKR}.  Furthermore, Fig. \ref{fig:LBskrvsN} illustrates how correlations between Eve and Bob channels degrade the overall SKR as \eqref{rkmlb} is a lowerbound of \eqref{rkmlbuncor} because of secrecy leakage and how the RIS significantly increases the SKR under certain scenarios}. Next, we study this impact in more detail.
\subsection{ Key Leakage}
{We can characterize the impact of channel correlations between Eve and Bob on the SKR. Assuming Eve is closer to Bob than Alice, we quantify the secrecy leakage incurred between Bob and Eve by finding the mutual information between $(\bar{G}_{\rm ab},\bar{G}_{\rm ae})$}.

\begin{corollary}
\label{corrSecrecy}
The secrecy leakage instigated by Eve's proximity to Bob characterized by the mutual information is 
\begin{align}
\label{Secrecy}
    I(\bar{G}_{\rm ab};\bar{G}_{\rm ae})&={\sf h}(\bar{G}_{\rm ab})+{\sf h}(\bar{G}_{\rm ae})-{\sf h}(\bar{G}_{\rm ab},\bar{G}_{\rm ae})\\
     &=-\log_2\bigg(1-\frac{(\rho_{ab}^{ae})^2}{\tilde{\rho}_{\rm ab}\tilde{\rho}_{\rm ae}}\bigg)\text{ bits.}
\end{align} 

\end{corollary}
\begin{IEEEproof}
{Using Lemma \ref{LemmaGaussian}, the channel estimates are complex Gaussian for large $N$. Then, we plug in the variances and the covariance matrix for $(\bar{G}_{\rm ab},\bar{G}_{\rm ae})$ given in \eqref{Covabe}, where ${\rm det}(\boldsymbol{\Sigma}_{abe})={\tilde{\rho}_{\rm ab}\tilde{\rho}_{\rm ae}}-(\rho_{ab}^{ae})^2$.}
\end{IEEEproof}
\begin{remark}
    Intuitively, in the uncorrelated case $
        I(\bar{G}_{\rm ab};\bar{G}_{\rm ae})=0$
    which corresponds to zero key leakage to Eve.   
\end{remark}

 


The RIS phases $\boldsymbol{\Phi}_\ell$ remain constant for $T_s$ symbols, which plays a role in the behavior of the estimation error contained in $\bar{\sigma}^2$. When $T_s$ increases the estimates improve and $\bar{\sigma}^2$ decreases.  Quantifying the effect of $T_s$ on the secrecy leakage will help determine appropriate values of $T_s$ during key generation. 
{The following corollary expresses this effect of $T_s$ on the secrecy leakage.}

\begin{corollary} The secrecy leakage satisfies
\begin{align}
  \lim_{T_s\to\infty}  I(\bar{G}_{\rm ab};\bar{G}_{\rm ae})=-\log_2\left(1-\frac{(\rho_{ab}^{ae})^2}{\rho_{\rm ab}\rho_{\rm ae}}\right) \text{ bits.}
\end{align}
  \end{corollary}
  \begin{IEEEproof}
{ Similar to Corollary \ref{corrSecrecy}, but the estimation error $\bar{\sigma}^2$ goes to zero in the limit. Thus,  $\tilde{\rho}_{\rm ab}=\rho_{\rm ab}$ and $\tilde{\rho}_{\rm ae}={\rho}_{\rm ae}$.}
  \end{IEEEproof}
   \begin{corollary}
Without the RIS, \eqref{Secrecy} becomes (as in \cite{pearson})
\begin{align}
    I(H_{\rm ab};H_{\rm ae})= -\log_2(1- \rho^2) \text{ bits.}
\end{align}
   \end{corollary}
   \begin{IEEEproof}
$I(H_{\rm ab};H_{\rm ae})={\sf h}(H_{\rm ab})+{\sf h}(H_{\rm ae})-{\sf h}(H_{\rm ab},H_{\rm ae})=\log(\pi e (\beta_{\rm ab}+\bar{\sigma}^2))+\log(\pi e (\beta_{\rm ae}+\bar{\sigma}^2))-\log(((\pi e)^2 (\beta_{\rm ae}+\bar{\sigma}^2)(\beta_{\rm ae}+\bar{\sigma}^2)(1-\rho^2))= -\log_2(1- \rho^2)$ bits.
   \end{IEEEproof}

{Although authors in \cite{otp} do not straightforwardly claim to have a separate CE phase in each transmission, they utilize the optimal RIS phase matrix in their information rate expression which depends on knowing the individual channels involved perfectly. On the other hand, the CE in our key generation phase {does not estimate individual channels but only the end-to-end channels $\bar{g}_{\rm ba}$ \eqref{Estimate_ba} and $\bar{g}_{\rm ab}$ \eqref{Estimate_ab}}. At a glance, having a dedicated and separate CE phase which estimates individual channels will improve the information rate, but it can also have negative effects on the system. The time reserved for the CE phase would degrade the overall secret information rate. {Furthermore, we show below that it can also lead to more secrecy leakage than a scheme without a dedicated CE  phase. }} After a dedicated CE phase, Eve obtains information on $h_{\rm ae}, \boldsymbol{h}_{\rm ar}, \boldsymbol{h}_{\rm re},\boldsymbol{h}_{\rm be},\boldsymbol{h}_{\rm br} $ in which the pairs $(h_{\rm ae},h_{\rm ab}) $ and $(\boldsymbol{h}_{\rm re},\boldsymbol{h}_{\rm rb})$ are highly spatially correlated given  Eve is close to Bob. {The following corollary shows that the secrecy leakage under a dedicated CE phase estimating individual channels is strictly larger than our scheme which only estimates the end-to-end channels.}
\begin{corollary}
{The secrecy leakage $ I(\bar{G}_{\rm ab};\bar{G}_{\rm ae},h_{\rm ae}, \boldsymbol{h}_{\rm 
    ar},\boldsymbol{h}_{\rm re}   ) $ instigated by having a dedicated CE phase to estimate individual channels $(h_{\rm ae}, \boldsymbol{h}_{\rm 
    ar},\boldsymbol{h}_{\rm re} )$ is lowerbounded as follows}
\begin{align}
\label{SecrecyCE}
    I(\bar{G}_{\rm ab};\bar{G}_{\rm ae},h_{\rm ae}, \boldsymbol{h}_{\rm 
    ar},\boldsymbol{h}_{\rm re}   ) >  I(\bar{G}_{\rm ab};\bar{G}_{\rm ae}).
\end{align} 
\end{corollary}
\begin{IEEEproof}
First, we can expand $ I(\bar{G}_{\rm ab};\bar{G}_{\rm ae},h_{\rm ae}, \boldsymbol{h}_{\rm 
    ar},\boldsymbol{h}_{\rm re}   )=I(\bar{G}_{\rm ab};\bar{G}_{\rm ae})+ I(\bar{G}_{\rm ab};h_{\rm ae}, \boldsymbol{h}_{\rm 
    ar},\boldsymbol{h}_{\rm re}|\bar{G}_{\rm ae})$ 
by the chain rule for mutual information.
Thus, it suffices to show that $I(\bar{G}_{\rm ab};h_{\rm ae}, \boldsymbol{h}_{\rm 
    ar},\boldsymbol{h}_{\rm re}|\bar{G}_{\rm ae})={\sf h}(\bar{G}_{\rm ab}|\bar{G}_{\rm ae})-{\sf h}(\bar{G}_{\rm ab}|h_{\rm ae}, \boldsymbol{h}_{\rm 
    ar},\boldsymbol{h}_{\rm re},\bar{G}_{\rm ae})$ is strictly positive, i.e. check
\begin{align}
        {\sf h}(\bar{G}_{\rm ab}|\bar{G}_{\rm ae})>{\sf h}(\bar{G}_{\rm ab}| h_{\rm ae},\boldsymbol{h}_{\rm 
    ar},\boldsymbol{h}_{\rm re},\bar{G}_{\rm ae}).
    \end{align}
  {Since conditioning reduces entropy, it suffices to show that }
    \begin{align}
        {\sf h}(\bar{G}_{\rm ab}|\bar{G}_{\rm ae})>{\sf h}(\bar{G}_{\rm ab}| h_{\rm ae},\bar{G}_{\rm ae}).
    \end{align}

{In closed form this is
  \begin{align} \log\left(\frac{\tilde{\rho}_{\rm ab}\tilde{\rho}_{\rm ae}-\tilde{\rho}_{\rm ab}^{{ae}^2}}{\tilde{\rho}_{\rm ae}}\right)>\log\left(\frac{\text{det}(\boldsymbol{E_2})}{\text{det}(\boldsymbol{E_1})}\right),
    \end{align}}
where
\begin{align}
\boldsymbol{E}_{1}&={\rm Cov}\left([ \bar{G}_{\rm ae}\ h_{ae} ]^T\right)=\begin{bmatrix}
\tilde{\rho}_{\rm ae} & \beta_{\rm ae} \\
 \beta_{\rm ae} &  \beta_{\rm ae}
\end{bmatrix},
\end{align}
and
\begin{align}
\boldsymbol{E}_{2}&={\rm Cov}\left([\bar{G}_{\rm ab}\ \bar{G}_{\rm ae} \ h_{ae} ]^T\right) \nonumber\\
&=\begin{bmatrix}
\tilde{\rho}_{\rm ab} & \tilde{\rho}_{\rm ab}^{ae}&  \rho\sqrt{\beta_{\rm ab}\beta_{\rm ae}} \\
\tilde{\rho}_{\rm ab}^{ae}& \tilde{\rho}_{\rm ae} &  \beta_{\rm ae}\\ \rho\sqrt{\beta_{\rm ab}\beta_{\rm ae}} &  \beta_{\rm ae}&  \beta_{\rm ae}
\end{bmatrix}.
\end{align}
Because of monotonocity of the logarithm, it is sufficient to check 
  \begin{align} 
\label{inequality}\frac{\tilde{\rho}_{\rm ab}\tilde{\rho}_{\rm ae}-\tilde{\rho}_{\rm ab}^{{\rm ae}^2}}{\tilde{\rho}_{\rm ae}}>\frac{\text{det}(\boldsymbol{E_2})}{\text{det}(\boldsymbol{E_1})}.
    \end{align}

  {With some work, and assuming without loss of generality that $\beta_{rb}=0$, we  can simplify \eqref{inequality} as
\begin{align}
 \frac{\beta_{\rm ab}((1-\rho^2)\beta_{\rm ae}+\beta_{\rm ar}\beta_{\rm re} \|\mathbf{R}\|_F^2)}{\beta_{\rm ae}+\beta_{\rm ar}\beta_{\rm re} \|\mathbf{R}\|_F^2}&>   (1-\rho^2)\beta_{\rm ab}.
 \end{align}
 This implies that 
 \begin{align}
\beta_{\rm ar}\beta_{\rm re} \|\mathbf{R}\|_F^2&>   (1-\rho^2)\beta_{\rm ar}\beta_{\rm re} \|\mathbf{R}\|_F^2\end{align}
and dividing both sides by $\beta_{\rm ar}\beta_{\rm re} \|\mathbf{R}\|_F^2$ finally yields 
 \begin{align}
1&>   (1-\rho^2),
\end{align}
which is true}, when $\rho$ is non-zero, i.e., Bob and Eve's channels are correlated. Therefore,  $  I(\bar{G}_{\rm ab};\bar{G}_{\rm ae},h_{\rm ae}, \boldsymbol{h}_{\rm 
    ar},\boldsymbol{h}_{\rm re}   ) >  I(\bar{G}_{\rm ab};\bar{G}_{\rm ae})$ which concludes the proof. 
\end{IEEEproof}

 This section discussed theoretical lower bounds for SKR under correlated and uncorrelated channels with and without RIS. It delved into quantifying key leakage and what parameters may affect it. We focus next on developing a practical scheme using insights from this section and present a novel protocol for secret key generation.

\section{Practical Implementation for Secret Key Generation}

In this section, we introduce a novel method for creating keys using several coherence blocks.
Over several blocks $f=1,\ldots, F$, we create $L=\frac{T_{\rm k}}{T_{\rm s}}$ keys, { each generated from $F$ estimates, using one estimate from each block.} Thus, in the $\ell^{th}$ switching period in $F$ blocks, Alice and Bob generate keys $\boldsymbol{k}_{\rm a, \ell}$ and $\boldsymbol{k}_{\rm b, \ell}$, respectively.  

{We describe the generation of these keys first in Sec. \ref{subQ}, describe the mismatch in key bits incurred in Sec. \ref{submismach}, followed by expressing the key rate that can be realized so that the two keys match in Sec. \ref{subeskr}. Finally, we analyze the probability that Eve intercepts the key in Sec. \ref{subEveleak}.} 
\subsection{Quantization of Channel Estimates} 
\label{subQ}
Following \cite{secureKeysMultipath}, Alice generates key bits from switching period $\ell$ in block $f$  by quantizing the phase of its channel estimate $\bar{g}_{\rm ba,\ell}$ given by
\begin{align}
    \theta_{\rm ba,\ell,f}&=\tan^{-1}\bigg(\frac{\text{imag}( \bar{g}_{\rm ba,\ell})}{\text{real}( \bar{g}_{\rm ba,\ell})}\bigg), \ell=1,\dots ,L.
\end{align}
Bob also does the same using the estimate $\bar{g}_{\rm ab,\ell}$. This happens over the several blocks $F$ until we obtain all the phases of all channel estimates.
We define the quantization of the phase using a function $f_Q:\mathbb{R} \rightarrow\{1,\dots, Q\}$, where $Q$ is the number of quantization levels, such that
\begin{align}
\label{quantizedphase}
 \theta_{ij,\ell, f}^Q=f_Q( \theta_{ij,\ell})=q, \text{ if } \theta_{ij,\ell,f}\in \left(\frac{2\pi(q-1)}{Q},\frac{2\pi(q)}{Q}\right),
 \end{align}
for $q=1,\ldots, Q$, $i\in\{\rm a,b\}$ with $i\neq j$. 

The total number of key bits in key $\ell$ (i.e., $\boldsymbol{k}_{\rm a,\ell}$ and $\boldsymbol{k}_{\rm b,\ell}$) denoted by $n$ is thus
\begin{align}
\label{keylength}
    n={F}\log_2(Q)\text{ bits}.
\end{align} 
A larger $Q$ increases the number of key bits at the expense of higher mismatch probability as discussed next.

 \subsection{Key Mismatch Rate}
  \label{submismach}
\begin{figure}
\centering
\tikzset{every picture/.style={scale=0.66}, every node/.style={scale=1}}
%
%
\begin{tikzpicture}

\begin{axis}[%
width=6.039in,
height=4.064in,
at={(1.013in,0.638in)},
scale only axis,
xmin=85,
xmax=110,
xlabel style={at={(0.7,0)},font=\color{white!15!black}},
xlabel={SNR (dB)},
xtick style={color=black},
xtick={85,90,95,100,105,110},
xticklabels={85,90,95,100,105,110},
ymin=0,
ymax=1,
ylabel style={at={(0,0.7)},font=\color{white!15!black}},
ylabel={$\text{Match Probability p}_\text{0}$},
axis background/.style={fill=white},
axis x line*=bottom,
axis y line*=left,
xmajorgrids,
ymajorgrids,
legend style={at={(0.847,0.204)}, anchor=south west, legend cell align=left, align=left, draw=white!15!black}
]
\addplot [color=black, line width=1.0pt]
  table[row sep=crcr]{%
85	0.767025886012665\\
90	0.848683479613376\\
95	0.918564603932896\\
100	0.951894233974003\\
105	0.973558493500722\\
110	0.984779468947895\\
};
\addlegendentry{Actual $p_0$, $Q=2$}

\addplot [color=black, line width=1.0pt, mark=o, mark options={solid, black}]
  table[row sep=crcr]{%
85	0.818447419029981\\
90	0.878615531831521\\
95	0.923625737158405\\
100	0.953974727302852\\
105	0.973031619575989\\
110	0.9844679165407\\
};
\addlegendentry{Approx. $\tilde{p}_0$, $Q=2$}

\addplot [color=blue, dashed, line width=1.0pt]
  table[row sep=crcr]{%
85	0.372625263859571\\
90	0.554938340184424\\
95	0.707476947005888\\
100	0.828352405288301\\
105	0.89667814687257\\
110	0.942895233862904\\
};
\addlegendentry{Actual $p_0$, $Q=8$}

\addplot [color=blue, line width=1.0pt, mark=o, mark options={solid, blue}]
  table[row sep=crcr]{%
85	0.257636435706868\\
90	0.486495159176632\\
95	0.695144514188473\\
100	0.829581970802756\\
105	0.90547550031066\\
110	0.947355627494189\\
};
\addlegendentry{Approx. $\tilde{p}_0$, $Q=8$}

\addplot [color=red, dashed, line width=1.0pt]
  table[row sep=crcr]{%
85	0.217642484168426\\
90	0.342295300522164\\
95	0.52771914231752\\
100	0.698700144428397\\
105	0.808910121097656\\
110	0.886012665259416\\
};
\addlegendentry{Actual $p_0$, $Q=16$}

\addplot [color=red, line width=1.0pt, mark=o, mark options={solid, red}]
  table[row sep=crcr]{%
85	0.079674893883364\\
90	0.207325039524113\\
95	0.424256413905044\\
100	0.648950832636802\\
105	0.803585268529934\\
110	0.891848520946598\\
};
\addlegendentry{Approx.  $\tilde{p}_0$, $Q=16$}

\end{axis}

\begin{axis}[%
width=7.792in,
height=5.083in,
at={(0in,0in)},
scale only axis,
xmin=0,
xmax=1,
ymin=0,
ymax=1,
axis line style={draw=none},
ticks=none,
axis x line*=bottom,
axis y line*=left
]
\end{axis}
\end{tikzpicture}%
    \caption{{Effect of SNR and $Q$ on the match probability. Parameters taken from Table \ref{tabsims}.}}
    \label{fig:matchq}. 
\end{figure}
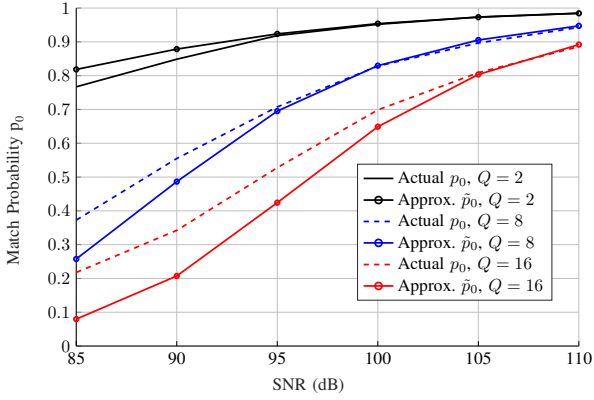

 {It is vital to quantify how the key bits for Alice and Bob match to characterize the overall secrecy rate and understand system performance. To this end, we quantify the key bits mismatch defined here as the key mismatch rate (KMR)}
\begin{align}
   p=   \frac{1}{FL}\sum_{f=1}^F\sum_{l=1}^LP(\theta_{\rm ba,\ell,f}^Q\neq\theta_{\rm ab,\ell,f}^Q),
\end{align} 
where $\theta_{\rm ba,\ell,f}^Q$ and $\theta_{\rm ab,\ell,f}^Q$ are as defined in \eqref{quantizedphase}. 
{In the limit as $F$ increases, we can use the law of large numbers to assert the approximation:
$p= \mathbb{E}[P(\theta_{\rm ba,\ell,f}^Q\neq\theta_{\rm ab,\ell,f}^Q)]$ averaging out the fluctuations incurred from the random RIS phase matrices and the varying channels which is useful in consequent analysis of key rate.}
Define $p_0=1-p$ as the match probability. A simplified approximation for $p_0$ is found in \cite[Proposition ~2]{secureKeysMultipath} which {is accurate} for high SNR. The approximation is 
\begin{align}
\label{probappprox}
 \tilde{p}_0=\frac{\tan^2(\frac{\pi}{Q})}{2(\tan^2(\frac{\pi}{Q})+\tilde{\sigma}^2)}+\frac{Q}{2\pi}\int_0^{\frac{\pi}{Q}}  \frac{\tan^2(\theta)}{\tan^2(\theta)+\tilde{\sigma}^2}d \theta
\end{align}
where $\tilde{\sigma}^2=\frac{1}{\mathbb{E}[\text{SNR}]}$.
{Fig. \ref{fig:matchq} shows the match probability versus SNR for different quantization levels. First, as the SNR of the received key bits increases so does the match probability. Second, when quantization levels increase, the match probability is diminished significantly because there is more chance for bit mismatches. Finally, it validates the approximation $ \tilde{p}_0$, which also provides insights on how it depends on SNR and $Q$.}

 \subsection{Effective Secret Key Rate}
 \label{subeskr}
 {We note that the quantity $\frac{2n}{F T_{\rm s}}$ where $n$ {is as given in }\eqref{keylength} represents a rate of generating key bits per transmission for each of Alice and Bob, but does not ensure that the keys match. To ensure that the keys match, we need a process that takes care of mismatched key bits, which will eventually diminish the SKR below $\frac{2n}{F T_{\rm s}}$. This is explained next. }

{Previously, we used channel quantization to obtain the keys $\boldsymbol{k}_{\rm a}, \boldsymbol{k}_{\rm b}$ at Alice and Bob, respectively. Using one time pad (OTP) which achieves perfect secrecy \cite{shannon}, Alice can encrypt message  $\boldsymbol{m}$ using $\boldsymbol{k}_{\rm a}$ as}
\begin{align}
\label{eqotp}
   \boldsymbol{c}=\boldsymbol{m} \oplus\boldsymbol{k}.
\end{align}
{
Bob  can apply $\boldsymbol{k}_{\rm b}$ to decrypt $\boldsymbol{c}$ but ends up with a corrupted message $\boldsymbol{m} +\boldsymbol{e} $ where $\boldsymbol{e} $ is a  binary error vector with $n(1-p_0)$ errors. This is indeed similar to a Binary Symmetric Channel (BSC) with a cross-over probability of $1-p_0$, {where the errors can be corrected} with a code with rate $(1 - {\sf H_b}(p_0))$ where ${\sf H_b}(\cdot)$ is the binary
entropy function.
}

{Thus, we can interpret this process as equivalent to OTP with matching keys with $n(1 - {\sf H_b}(p_0))$ key bits leading to the effective SKR. Using this reasoning, we can conclude with the following theorem.}
\begin{theorem}
   The effective secret key rate (ESKR) $    \bar{R}_{\rm k}$ for OTP with mismatched keys is characterized as 
\begin{align}
\label{avgthrough}
    \bar{R}_{\rm k}=\frac{(1 - {\sf H_b}(p_0))\log_2(Q)}{T_{\rm s}/2} \text{ bits per symbol.}
\end{align}
\end{theorem}

{To sum up, one can use the mismatched keys in transmission, and correct the errors introduced by the mismatched keys after decryption using a channel code for a $BSC(1-p_0)$.}

\begin{figure}
\centering
\tikzset{every picture/.style={line width=0.7pt}} 

\begin{tikzpicture}[x=0.53pt,y=0.6pt,yscale=-1,xscale=1]

\draw    (140,69) -- (182,69) ;
\draw [shift={(184,69)}, rotate = 180] [color={rgb, 255:red, 0; green, 0; blue, 0 }  ][line width=0.75]    (10.93,-3.29) .. controls (6.95,-1.4) and (3.31,-0.3) .. (0,0) .. controls (3.31,0.3) and (6.95,1.4) .. (10.93,3.29)   ;
\draw    (238,69) -- (279,69) ;
\draw [shift={(281,69)}, rotate = 180] [color={rgb, 255:red, 0; green, 0; blue, 0 }  ][line width=0.75]    (10.93,-3.29) .. controls (6.95,-1.4) and (3.31,-0.3) .. (0,0) .. controls (3.31,0.3) and (6.95,1.4) .. (10.93,3.29)   ;
\draw    (360,69) -- (401,69) ;
\draw [shift={(403,69)}, rotate = 180] [color={rgb, 255:red, 0; green, 0; blue, 0 }  ][line width=0.75]    (10.93,-3.29) .. controls (6.95,-1.4) and (3.31,-0.3) .. (0,0) .. controls (3.31,0.3) and (6.95,1.4) .. (10.93,3.29)   ;
\draw    (325,34) -- (324.09,55) ;
\draw [shift={(324,57)}, rotate = 272.49] [color={rgb, 255:red, 0; green, 0; blue, 0 }  ][line width=0.75]    (10.93,-3.29) .. controls (6.95,-1.4) and (3.31,-0.3) .. (0,0) .. controls (3.31,0.3) and (6.95,1.4) .. (10.93,3.29)   ;
\draw    (455,70) -- (488,70) -- (496,70) ;
\draw [shift={(498,70)}, rotate = 180] [color={rgb, 255:red, 0; green, 0; blue, 0 }  ][line width=0.75]    (10.93,-3.29) .. controls (6.95,-1.4) and (3.31,-0.3) .. (0,0) .. controls (3.31,0.3) and (6.95,1.4) .. (10.93,3.29)   ;
\draw  [fill={rgb, 255:red, 79; green, 169; blue, 186 }  ,fill opacity=1 ] (502.73,61.47) .. controls (502.14,57.56) and (504.09,53.69) .. (507.77,51.5) .. controls (511.45,49.31) and (516.21,49.18) .. (520.02,51.18) .. controls (521.37,48.91) and (523.84,47.34) .. (526.69,46.95) .. controls (529.54,46.55) and (532.43,47.39) .. (534.48,49.19) .. controls (535.63,47.13) and (537.89,45.75) .. (540.46,45.53) .. controls (543.02,45.31) and (545.53,46.29) .. (547.1,48.13) .. controls (549.18,45.94) and (552.49,45.02) .. (555.59,45.76) .. controls (558.7,46.51) and (561.04,48.78) .. (561.61,51.6) .. controls (564.16,52.22) and (566.28,53.8) .. (567.43,55.92) .. controls (568.58,58.05) and (568.64,60.52) .. (567.6,62.69) .. controls (570.11,65.61) and (570.7,69.49) .. (569.14,72.9) .. controls (567.59,76.3) and (564.13,78.72) .. (560.05,79.24) .. controls (560.02,82.43) and (558.06,85.37) .. (554.92,86.9) .. controls (551.78,88.44) and (547.95,88.35) .. (544.91,86.65) .. controls (543.62,90.48) and (539.98,93.3) .. (535.56,93.89) .. controls (531.14,94.48) and (526.74,92.73) .. (524.26,89.41) .. controls (521.21,91.04) and (517.56,91.52) .. (514.13,90.72) .. controls (510.69,89.91) and (507.76,87.91) .. (505.99,85.15) .. controls (502.88,85.47) and (499.88,84.03) .. (498.47,81.54) .. controls (497.05,79.06) and (497.54,76.05) .. (499.68,74.01) .. controls (496.9,72.55) and (495.49,69.66) .. (496.17,66.84) .. controls (496.85,64.02) and (499.47,61.92) .. (502.67,61.62) ; \draw   (499.68,74.01) .. controls (500.99,74.7) and (502.5,75.01) .. (504.01,74.91)(505.99,85.15) .. controls (506.64,85.08) and (507.28,84.94) .. (507.89,84.72)(524.26,89.41) .. controls (523.8,88.79) and (523.41,88.14) .. (523.11,87.45)(544.91,86.65) .. controls (545.15,85.96) and (545.3,85.24) .. (545.37,84.51)(560.05,79.24) .. controls (560.08,75.83) and (557.92,72.72) .. (554.49,71.23)(567.6,62.69) .. controls (567.05,63.85) and (566.2,64.88) .. (565.12,65.69)(561.61,51.6) .. controls (561.71,52.07) and (561.75,52.54) .. (561.74,53.02)(547.1,48.13) .. controls (546.58,48.67) and (546.15,49.28) .. (545.83,49.93)(534.48,49.19) .. controls (534.2,49.69) and (534,50.21) .. (533.86,50.75)(520.02,51.18) .. controls (520.83,51.6) and (521.57,52.11) .. (522.24,52.69)(502.73,61.47) .. controls (502.82,62.01) and (502.95,62.54) .. (503.12,63.06) ;
\draw    (185,148.5) -- (143,148.5) ;
\draw [shift={(141,148.5)}, rotate = 360] [color={rgb, 255:red, 0; green, 0; blue, 0 }  ][line width=0.75]    (10.93,-3.29) .. controls (6.95,-1.4) and (3.31,-0.3) .. (0,0) .. controls (3.31,0.3) and (6.95,1.4) .. (10.93,3.29)   ;
\draw    (284,148.5) -- (242,148.5) ;
\draw [shift={(240,148.5)}, rotate = 360] [color={rgb, 255:red, 0; green, 0; blue, 0 }  ][line width=0.75]    (10.93,-3.29) .. controls (6.95,-1.4) and (3.31,-0.3) .. (0,0) .. controls (3.31,0.3) and (6.95,1.4) .. (10.93,3.29)   ;
\draw    (406,149.5) -- (364,149.5) ;
\draw [shift={(362,149.5)}, rotate = 360] [color={rgb, 255:red, 0; green, 0; blue, 0 }  ][line width=0.75]    (10.93,-3.29) .. controls (6.95,-1.4) and (3.31,-0.3) .. (0,0) .. controls (3.31,0.3) and (6.95,1.4) .. (10.93,3.29)   ;
\draw    (527,92.5) -- (528,149.5) ;
\draw    (528,149.5) -- (463,149.5) ;
\draw [shift={(461,149.5)}, rotate = 360] [color={rgb, 255:red, 0; green, 0; blue, 0 }  ][line width=0.75]    (10.93,-3.29) .. controls (6.95,-1.4) and (3.31,-0.3) .. (0,0) .. controls (3.31,0.3) and (6.95,1.4) .. (10.93,3.29)   ;
\draw    (323,190.5) -- (323,170.5) ;
\draw [shift={(323,168.5)}, rotate = 90] [color={rgb, 255:red, 0; green, 0; blue, 0 }  ][line width=0.75]    (10.93,-3.29) .. controls (6.95,-1.4) and (3.31,-0.3) .. (0,0) .. controls (3.31,0.3) and (6.95,1.4) .. (10.93,3.29)   ;

\draw    (184.5,53) -- (237.5,53) -- (237.5,91) -- (184.5,91) -- cycle  ;
\draw (211,72) node  [font=\scriptsize] [align=left] {Outer \\Encoder};
\draw (118,66) node  [font=\footnotesize] [align=left] {Alice};
\draw    (402.5,54) -- (455.5,54) -- (455.5,92) -- (402.5,92) -- cycle  ;
\draw (429,73) node  [font=\scriptsize] [align=left] {Inner\\Encoder};
\draw    (283,56) -- (360,56) -- (360,82) -- (283,82) -- cycle  ;
\draw (286,60.4) node [anchor=north west][inner sep=0.75pt] [font=\scriptsize]   {$\boldsymbol{s} =\boldsymbol{b}\oplus \boldsymbol{k}_{a}$};
\draw (320,15.4) node [anchor=north west][inner sep=0.75pt]    {$\boldsymbol{k}_{a}$};
\draw (254,73.4) node [anchor=north west][inner sep=0.75pt]    {$\boldsymbol{b}$};
\draw (375,72.4) node [anchor=north west][inner sep=0.75pt]    {$\boldsymbol{s}$};
\draw (474,72.4) node [anchor=north west][inner sep=0.75pt]    {$\boldsymbol{x}$};
\draw (534,67) node  [font=\footnotesize] [align=left] {Channel};
\draw    (406.5,131) -- (459.5,131) -- (459.5,169) -- (406.5,169) -- cycle  ;
\draw (433,150) node  [font=\scriptsize] [align=left] {Outer\\Decoder};
\draw    (285,136) -- (362,136) -- (362,168) -- (285,168) -- cycle  ;
\draw (288,140.4) node [anchor=north west][inner sep=0.75pt]   [font=\scriptsize]  {$\hat{\boldsymbol{s}} =\hat{\boldsymbol{b}}\oplus\boldsymbol{ k}_{b}$};
\draw    (185.5,131) -- (238.5,131) -- (238.5,169) -- (185.5,169) -- cycle  ;
\draw (212,150) node  [font=\scriptsize] [align=left] {Inner\\Decoder};
\draw (119,146) node  [font=\footnotesize] [align=left] {Bob};
\draw (479.5,151.4) node [anchor=north west][inner sep=0.75pt]    {$\boldsymbol{y}$};
\draw (376,151.9) node [anchor=north west][inner sep=0.75pt]    {$\hat{\boldsymbol{s}}$};
\draw (255,147.4) node [anchor=north west][inner sep=0.75pt]    {$\hat{\boldsymbol{b}}$};
\draw (315,190.4) node [anchor=north west][inner sep=0.75pt]    {$\boldsymbol{k}_{b}$};
\draw (154,73.4) node [anchor=north west][inner sep=0.75pt]    {$\boldsymbol{m}$};
\draw (158,150.4) node [anchor=north west][inner sep=0.75pt]    {$\boldsymbol{m}$};

\end{tikzpicture}
    \caption{Schematic of achievable secret transmission using dual-stage encoding and OTP encryption.}
\label{fig:aliceandbobschematic}
\end{figure}
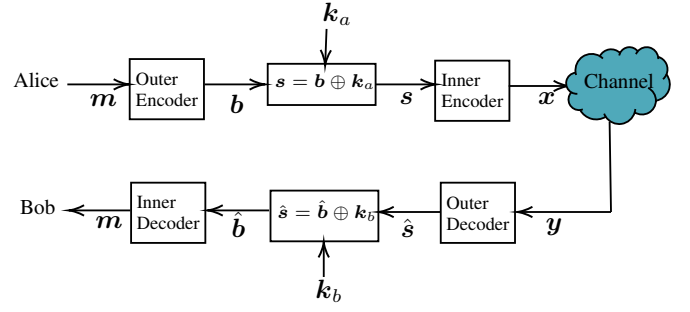
{For clarity,  Fig. \ref{fig:aliceandbobschematic} shows Alice's message $\boldsymbol{m}$ with $m$ bits is encoded by the outer encoder into $\boldsymbol{b}$ with $n$ bits to remove key errors (via information reconciliation phase), where $n>m$, then passed into an encrypting function with the key $\boldsymbol{k}_a$ to create encrypted message $\boldsymbol{s}$. Finally, it is encoded to remove channel errors by the inner encoder into $\boldsymbol{x}$ which is then sent over the wireless medium. At Bob's side, the erroneous message $\boldsymbol{y}$ first passes through the outer decoder to remove any errors resulting from the channel to get $\hat{\boldsymbol{s}}$, which is then passed into the decrypting function with Bob's likely mismatched $\boldsymbol{k}_b$ to produce $\hat{\boldsymbol{b}}$. Finally, it passes through the inner decoder which removes errors incurred from key mismatches to get the original intended message $\boldsymbol{m}$ given the decoding and encoding process eliminated all possible errors (from channel or key generation). Effectively, this creates a matched shorter key in which to encrypt with a shorter message.}

\begin{figure}[t]
\tikzset{every picture/.style={scale=0.68}, every node/.style={scale=  1}}
%
%
\begin{tikzpicture}

\begin{axis}[%
width=6.028in,
height=4.751in,
at={(1.011in,0.644in)},
scale only axis,
xmin=0,
xmax=25,
xlabel style={at={(0.623,0)},font=\color{white!15!black}},
xlabel={$L$ RIS Switching Intervals},
ymin=0.051,
ymax=0.435,
yminorticks=true,
ylabel style={at={(0,0.6)},font=\color{white!15!black}},
ylabel={Key Mismatch Rate $p$},
axis background/.style={fill=white},
xmajorgrids,
ymajorgrids,
yminorgrids,
legend style={at={(0.623,0.145)}, anchor=south west, legend cell align=left, align=left, draw=white!15!black}
]
\addplot [color=black, line width=1.0pt]
  table[row sep=crcr]{%
25	0.2468\\
12.5	0.1816\\
8.33333333333333	0.1456\\
6.25	0.1368\\
5	0.1234\\
4.16666666666667	0.1084\\
3.57142857142857	0.106\\
3.125	0.089\\
2.77777777777778	0.097\\
2.5	0.091\\
2.27272727272727	0.084\\
2.08333333333333	0.0742\\
1.92307692307692	0.073\\
1.78571428571429	0.0654\\
1.66666666666667	0.0648\\
1.5625	0.066\\
1.47058823529412	0.0684\\
1.38888888888889	0.0594\\
1.31578947368421	0.0646\\
1.25	0.0596\\
1.19047619047619	0.0584\\
1.13636363636364	0.051\\
1.08695652173913	0.06\\
1.04166666666667	0.0538\\
1	0.0592\\
};
\addlegendentry{Between Bob and Alice}

\addplot [color=red, line width=1.0pt]
  table[row sep=crcr]{%
25	0.4202\\
12.5	0.4222\\
8.33333333333333	0.4124\\
6.25	0.4038\\
5	0.4188\\
4.16666666666667	0.4076\\
3.57142857142857	0.418\\
3.125	0.416\\
2.77777777777778	0.4022\\
2.5	0.4042\\
2.27272727272727	0.4184\\
2.08333333333333	0.401\\
1.92307692307692	0.417\\
1.78571428571429	0.4086\\
1.66666666666667	0.3956\\
1.5625	0.4052\\
1.47058823529412	0.3998\\
1.38888888888889	0.3994\\
1.31578947368421	0.409\\
1.25	0.4008\\
1.19047619047619	0.4074\\
1.13636363636364	0.411\\
1.08695652173913	0.3988\\
1.04166666666667	0.4032\\
1	0.4124\\
};
\addlegendentry{Between Eve and Alice}

\addplot [color=blue, line width=1.0pt, only marks, mark=o, mark options={solid, blue}]
  table[row sep=crcr]{%
25	0.435\\
12.5	0.415\\
8.33333333333333	0.41\\
6.25	0.4058\\
5	0.4106\\
4.16666666666667	0.4116\\
3.57142857142857	0.4176\\
3.125	0.4126\\
2.77777777777778	0.4024\\
2.5	0.4124\\
2.27272727272727	0.412\\
2.08333333333333	0.4076\\
1.92307692307692	0.4124\\
1.78571428571429	0.4128\\
1.66666666666667	0.3948\\
1.5625	0.3992\\
1.47058823529412	0.395\\
1.38888888888889	0.4016\\
1.31578947368421	0.4044\\
1.25	0.3996\\
1.19047619047619	0.411\\
1.13636363636364	0.4112\\
1.08695652173913	0.4008\\
1.04166666666667	0.3982\\
1	0.416\\
};
\addlegendentry{Between Eve and Bob}

\end{axis}

\begin{axis}[%
width=7.778in,
height=5.833in,
at={(0in,0in)},
scale only axis,
xmin=0,
xmax=1,
ymin=0,
ymax=1,
axis line style={draw=none},
ticks=none,
axis x line*=bottom,
axis y line*=left
]
\end{axis}
\end{tikzpicture}%
\caption{{KMR versus the number of RIS switching intervals $L$ for ALice and Bob, Alice and Eve, and Bob and Eve keys, respectively. Parameters taken from Table \ref{tabsims}, and Eve’s position
is at $(31, 0, 1.5)m$.}}
    \label{fig:keymis}
\end{figure}
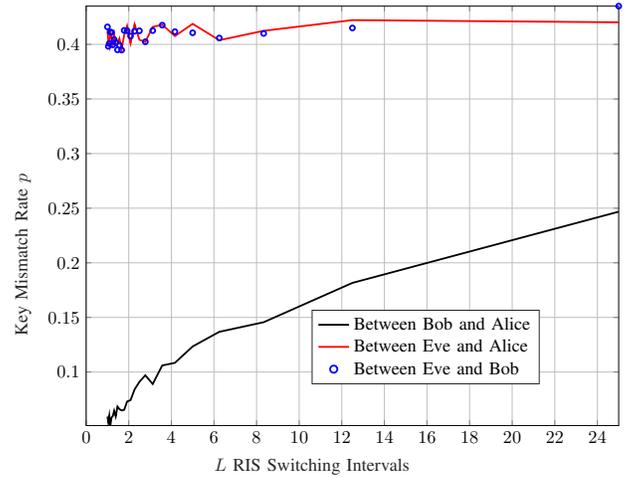


{Fig. \ref{fig:keymis} shows the {increase of the }mismatch probability between Alice and Bob keys  against $L$. This is because as $L$ increases the channel estimates observed at Alice and Bob, respectively, incur more errors since $T_{\rm s}$ (generally the channel estimation time for the symbol) decreases. Furthermore, it demonstrates that a channel code can in theory correct key mismatched induced errors between Alice and Bob, however, Eve cannot do the same due to the large number of errors.
}

\subsection{ Eve Intercept Probability }
\label{subEveleak}
{Previously, we characterized the SKR and extracted key $\boldsymbol{k}$ at Alice and Bob. Now, we quantify the secrecy leakage incurred from a successful eavesdropping event from Eve.}

\begin{lemma}
 Let $E$ be the event in which Eve guesses the {entire} key, then the probability $P(E)$ is upper-bounded by \cite{pearson}
 \begin{align}
 \label{probEve}
         P(E) &\leq  \left(\sqrt{2I (\bar{G}_{\rm ab};\bar{G}_{\rm ae})}+\frac{1}{Q}\right)^F
 \end{align}
\end{lemma}
\begin{IEEEproof}
    First, the probability of Eve guessing {$\log_2(Q)$ key bits generated from one channel estimate} correctly given no observations is upper-bounded by $\frac{1}{Q}$. Furthermore, given observations, this probability is increased by at most $\sqrt{2I (\bar{G}_{\rm ab};\bar{G}_{\rm ae})}$ \cite{pearson}. Over the number of channels uses to create the key $F$, this becomes as \eqref{probEve}.
\end{IEEEproof}

{Next, it is imperative to formulate the secrecy  rate.}


\section{Secrecy Rate}
We define the secrecy rate as the number of secret bits sent divided by the total time for the transmission. The number of secret bits is the number of information bits successfully encrypted in \eqref{eqotp} which is the minimum length of either the established key or the message.  { The length of the message depends on the achievable rate of the transmission scheme during the information transmission phase, which we investigate next. }

The secrecy rate $ R_{\rm S}$ is determined by taking the minimum of the ergodic information rate and the key rate, i.e.,
\begin{align}
\label{secrecycap}
    R_{\rm S}=\text{min}\left\{\frac{T_{\rm k}}{2T}\bar{R}_{\rm k}, \left(\frac{T-T_{\rm k}}{T}\right)  R_{\rm I}\right\},
\end{align}
where $\bar{R}_{\rm k}$ is defined in \eqref{avgthrough} and 
\begin{align}
R_{\rm I}&=\mathbb{E}[\log(1+\gamma(\boldsymbol{\Phi}_{l^*}))]
\label{inforateerogdic}
\end{align}
is
 the ergodic information rate, where $\gamma(\boldsymbol{\Phi}_{l^*})$ is the random SNR from Alice to Bob link, and the expectation is taken over  the channel samples and RIS configurations in different consecutive frames. {The ergodic information rate is characterized next.}

\subsection{Secure Information Rate}
During the first $T_{\rm k}$ symbols, $\boldsymbol{\Phi}_l$ is chosen arbitrarily at random every switching interval $\ell$. This ensures channel disruption during channel probing to create a lengthier key. During the transmission phase, however,  $\boldsymbol{\Phi}_l$ remains fixed. 
To maximize the achievable information rate, we can rely on the CSI gained via the key generation phase and utilize opportunistic beamforming (OB) via the RIS to choose the maximum SNR value.  
In particular, during the key generation phase, multiple RIS configurations $\boldsymbol{\Phi}_{l}, l=1,\dots,L$ are used and  {Bob uses its channel estimates to find the index $l^*$ }and send it to
Alice over a public channel\footnote{{Alternatively, because of reciprocity Alice can estimate the SNR use it as the actual SNR from Alice to Bob which lowers latency.}} where 
\begin{align}
\label{eqRISmax}
 l^*=\underset{l=1,3,5\dots L}{\text{argmax}}\hspace{0.1in} |\bar{g}_{\rm ab,\ell}|. 
\end{align}
 {The RIS can then deploy $\boldsymbol{\Phi}_{l^*}$ for the transmission phase in that frame.}
 In short, we {deploy} the RIS phase shift matrix that ensures a best overall channel. If $L$ is large enough given $N$, i.e., the RIS changes its configuration frequently enough in relevance to its number of elements, then  it becomes more likely to choose a pseudo-random configuration close to the optimal configuration for the given channel realization. Intuitively, this allows the system  to effectively ride the peaks of the fading channels \cite{Nadeem2021OpportunisticBU,opportunisticBS}.

To investigate the impact of the RIS on the information rate analysis, we need the following theorem.
\begin{theorem}
\label{corChannelDist}
    Under random RIS rotations and with $N$ large, the overall channel $\bar{g}_{\rm ab,\ell}=h_{\rm ab} + \boldsymbol{h}_{\rm ar}^H \boldsymbol{\Phi}_{\ell} \boldsymbol{h}_{\rm rb}$ represented by random variable $G_{\rm ab}$ can be modeled as 
    \begin{align}
          {G}_{\rm ab} \sim \mathcal{CN}(0,\beta_{\rm ab} + \beta_{ar}\beta_{rb}\|\mathbf{R}\|_F^2).
    \end{align}
\end{theorem}
\begin{IEEEproof}
The proof is found in \cite{impactofPhaseRandomIRS}.
\end{IEEEproof}
The following theorem analyses \eqref{avgthrough} in closed form and gives insight on how RIS switching rate $L$ affects the ergodic rate.
\begin{theorem}
Under Theorem \ref{corChannelDist}, the ergodic rate in \eqref{inforateerogdic} during downlink secure information transmission scales for $L$ large as
\begin{align}
\label{eqravgscale}
R_{\rm I}^{(L)}=\log\bigg(\frac{\log (L)(\beta_{\rm ab} + \beta_{ar}\beta_{rb}\|\mathbf{R}\|_F^2)P}{\sigma^2}\bigg)+o(1), 
\end{align}
{where $o(1)$ represents the terms that go to zero as $L \rightarrow \infty$}.
\end{theorem}
\begin{IEEEproof}
    The channel $\bar{g}_{\rm ab,\ell}$ is statistically equivalent to $ \sqrt{\beta} \tilde{h}_l$ where $\beta =\beta_{\rm ab} +\beta_{ar}\beta_{rb}\|\mathbf{R}\|_F^2$ and $\tilde{h}_l \sim \mathcal{CN}(0,1)$.
For large $L$, the behavior of $\underset{l}{\text{max}}|\tilde{h}_l|^2$ can be modelled as 
\begin{align}
    \underset{l}{\text{max}}|\tilde{h}_l|^2\triangleq \log L+\mathcal{O}(\log\log\log L),
\end{align}
using the Extreme Value Theorem \cite{randomrot,scalingLaw} since ${\text{max}}|\tilde{h}_l|^2$ is the maximum of i.i.d. Chi-square RVs. Thus, we can write
    \begin{align}
R_{\rm I}^{(L)}&=
\log\left(1+\frac{\beta P}{\sigma^2}\log L+\mathcal{O}(\log\log\log L)\right)\\&=\log\bigg(\frac{\beta P}{\sigma^2}\log L\bigg[1+\frac{\sigma^2}{\beta P\log L}+\mathcal{O}\bigg(\frac{\log\log\log L}{\log L}\bigg)\bigg]\bigg)\\&=\log\bigg(\frac{\log (L)(\beta_{\rm ab} + \beta_{ar}\beta_{rb}\|\mathbf{R}\|_F^2)P}{\sigma^2}\bigg)+o(1),
\end{align} for $L$ large.
\end{IEEEproof}
{Next, we present some useful benchmarks for the ergodic information rate.}

\subsection{Benchmarks}

{ {Assuming perfect CSI}, we can attain the optimal RIS matrix $ \boldsymbol{\Phi}^*$ 
 in which the phases of the diagonal components are defined as \cite{nadeem2020intelligent}}
\begin{align}
\label{risaligned}
    \angle {\Phi}^*_{n,n}=\angle h_{ar,n}-\angle h_{rb,n}+\angle h_{\rm ab}.
    \end{align}
{ To compare with \cite{otp},  we evaluate the information rate under \eqref{risaligned} next.\footnote{Note that perfect CSI is often an idealistic assumption and cannot be applied in practice.}} The authors in \cite{otp} implicitly make the assumption that perfect CSI is available  at either Alice or Bob, thus they can obtain the optimal RIS matrix in \eqref{risaligned}. However, this is idealistic, impractical and can be detrimental to the secrecy rate since it needs long coherence interval for the extensive channel estimation which scales with $N$, and structured schemes in key generation scenarios can pose as a security risk.

{ We compare with \cite{otp} by considering an upper-bound on the information rate under optimal RIS phase shift which is defined next and will be used as a meaningful benchmark in subsequent simulations.}
\begin{corollary}Under perfect CSI,  the optimal  RIS configuration $ \boldsymbol{\Phi}^*$  can be used to optimize the information rate $R_{\rm I}^{*}$ as
 \begin{align}
\label{eqravgoptim}
R_{\rm I}^*&=\mathbb{E}\left[\log\left(1+\frac{P|h_{\rm ab} + \boldsymbol{h}_{\rm ar}^H \boldsymbol{\Phi}^* \boldsymbol{h}_{\rm rb}  |^2}{\sigma^2}\right)\right]\\&{=}\mathbb{E}\left[\log\left(1+\frac{P(|h_{\rm ab}| + \sum_{n=1}^N|{h}_{\rm ar,n}^H|| {h}_{\rm rb,n}  |)^2}{\sigma^2}\right)\right].
\end{align}
\end{corollary}


The secrecy rate $ R_{\rm S}^{*}$ is then determined by taking the minimum of the optimized information rate rate and the ESKR, i.e.,
\begin{align}
\label{secrecycap}
    R_{\rm S}^{*}=\text{min}\left\{\frac{T_{\rm k}}{2T}\bar{R}_{\rm k}, \left(\frac{T-T_{\rm k}}{T}\right)  R_{\rm I}^{*}\right\},
\end{align}
\begin{corollary}
 Under the influence of only the direct channel, the average rate denoted by $  R_{\rm I}^{D}$ is 
 \begin{align}
 \label{eqravgnr}
 R_{\rm I}^{\rm D}&=\mathbb{E}\left[\log\left(1+\frac{|h_{\rm ab}|^2P}{\sigma^2}\right)\right], 
 \end{align}
\end{corollary} 
{which will be evaluated numerically in the simulations.}
The secrecy rate $ R_{\rm S}^{\rm D}$ is then determined by taking the minimum of the ergodic rate and the direct channel key rate, i.e.,
\begin{align}
\label{secrecyd1}
    R_{\rm S}^{\rm D 1}=\text{min}\left\{\frac{T_{\rm k}}{2T}\bar{R}_{\rm k}^{\rm D}, \left(\frac{T-T_{\rm k}}{T}\right)  R_{\rm I}^{\rm D}\right\},
\end{align}
in which $\bar{R}_{\rm k}^{\rm D}=\frac{(1-\sf{h}(p))\log_2(Q)}{T_{\rm k}/2}$.

In the case of a key-less based scenario, we can have a wiretap channel between Alice and Bob similar to works in \cite{wiretapWyner,wiretap1,wiretap2}. We formulate the secrecy rate under a key-less scenario in the next lemma.
 
\begin{lemma} {Under the direct channel {only, i.e. without an RIS,} the secrecy rate
for key-less secrecy is given by}
    \begin{align}
\label{secrecyd2}
    R_{\rm S}^{\rm D2}&=
 \left( \mathbb{E} \left[\log\left(1+\frac{|h_{\rm ab}|^2P}{\sigma^2}\right)\right]-\mathbb{E} \left[\log\left(1+\frac{|h_{\rm ae}|^2P}{\sigma^2}\right)\right]\right)^+ 
\end{align}
\end{lemma}
where $\left(x \right)^+$
 denotes the $max(0, x)$ and $i$ refers to Jensen's inequality.
\begin{IEEEproof}
    The proof can be found in \cite{wiretapWyner}.
\end{IEEEproof}

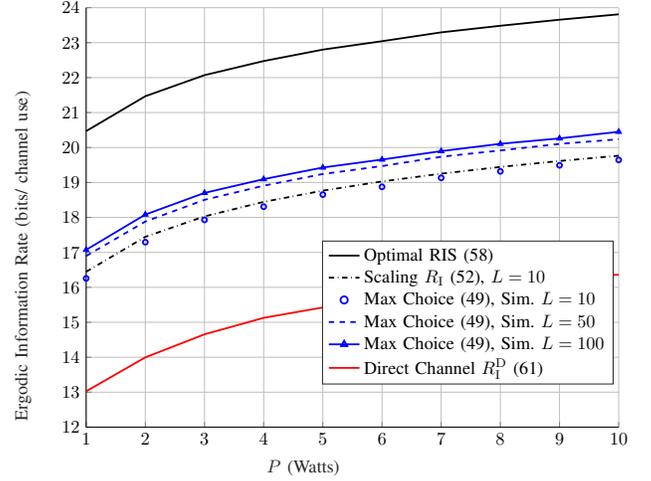
\begin{figure}[t]
\tikzset{every picture/.style={scale=0.68}, every node/.style={scale=1}}
%
%
\begin{tikzpicture}

\begin{axis}[%
width=6.028in,
height=4.751in,
at={(1.011in,0.644in)},
scale only axis,
xmin=1,
xmax=10,
xlabel style={at={(0.6,0)},font=\color{white!15!black}},
xlabel={$P$ (Watts)},
ymin=12,
ymax=24,
ylabel style={at={(0,0.6)},font=\color{white!15!black}},
ylabel={Ergodic Information Rate  (bits/ channel use)},
axis background/.style={fill=white},
axis x line*=bottom,
axis y line*=left,
xmajorgrids,
ymajorgrids,
legend style={at={(0.6518,0.15)}, anchor=south west, legend cell align=left, align=left, draw=white!15!black}
]
\addplot [color=black, line width=1.0pt]
  table[row sep=crcr]{%
1	20.468753522286\\
2	21.467508066666\\
3	22.0704661487177\\
4	22.4738608810064\\
5	22.8010117189433\\
6	23.0421006564533\\
7	23.2947187462906\\
8	23.4835982456916\\
9	23.6568737544692\\
10	23.8112238268622\\
};
\addlegendentry{Optimal RIS \eqref{eqravgoptim}}

\addplot [color=black, dashdotted, line width=1.0pt]
  table[row sep=crcr]{%
1	16.4444951405269\\
2	17.4444951405269\\
3	18.0294576412481\\
4	18.4444951405269\\
5	18.7664232354143\\
6	19.0294576412481\\
7	19.2518500625845\\
8	19.4444951405269\\
9	19.6144201419692\\
10	19.7664232354143\\
};
\addlegendentry{ Scaling $R_{\rm I}$ \eqref{eqravgscale},  $L=10$}

\addplot [color=blue, line width=1.0pt, only marks, mark=o, mark options={solid, blue}]
  table[row sep=crcr]{%
1	16.2567475178235\\
2	17.2922379989906\\
3	17.9302378974088\\
4	18.3070423033632\\
5	18.6532145347208\\
6	18.8754057192097\\
7	19.1322141865975\\
8	19.3174829628729\\
9	19.4914212262884\\
10	19.6446876567811\\
};
\addlegendentry{Max Choice \eqref{inforateerogdic}, Sim. $L=10$}

\addplot [color=blue, dashed, line width=1.0pt]
  table[row sep=crcr]{%
1	16.8957040809704\\
2	17.8820347224149\\
3	18.5027017777396\\
4	18.907304040968\\
5	19.2419239688588\\
6	19.4689825122645\\
7	19.7359566807356\\
8	19.9162187865867\\
9	20.1032291288726\\
10	20.237235351424\\
};
\addlegendentry{Max Choice \eqref{inforateerogdic}, Sim. $L=50$}

\addplot [color=blue, line width=1.0pt, mark=triangle, mark options={solid, blue}]
  table[row sep=crcr]{%
1	17.0728005026992\\
2	18.0814032690579\\
3	18.7011398381549\\
4	19.095262326873\\
5	19.4274268874363\\
6	19.6556294605594\\
7	19.8966791967518\\
8	20.1064744286588\\
9	20.261023434802\\
10	20.4501428739095\\
};
\addlegendentry{Max Choice \eqref{inforateerogdic}, Sim. $L=100$}

\addplot [color=red, line width=1.0pt]
  table[row sep=crcr]{%
1	13.0247207146101\\
2	13.995733829769\\
3	14.6548885655553\\
4	15.1265133553954\\
5	15.4236914856299\\
6	15.6039791947132\\
7	15.8895066056411\\
8	16.0720523155689\\
9	16.2700042601948\\
10	16.3611402085231\\
};
\addlegendentry{Direct Channel $R_{\rm I}^{\rm D}$ \eqref{eqravgnr}}

\end{axis}

\begin{axis}[%
width=7.778in,
height=5.833in,
at={(0in,0in)},
scale only axis,
xmin=0,
xmax=1,
ymin=0,
ymax=1,
axis line style={draw=none},
ticks=none,
axis x line*=bottom,
axis y line*=left
]
\end{axis}
\end{tikzpicture}%
\caption{{Ergodic information rate $R_{\rm I}$ versus $SNR=\frac{P}{\sigma^2}$ and benchmarks for different values of $L$. Simulation  parameters are found in Table \ref{tabsims}.}}
    \label{fig:inforate}
\end{figure}

{Fig. \ref{fig:inforate} highlights the performance gained when deploying the RIS in the information rate and validates the scaling behavior in \eqref{eqravgscale}. The figure shows how the gap between the information rate under optimal RIS and choosing the best configuration out of a set of $L$ configurations shrinks as $L$ increases.}

 After formulating the secrecy rate in \eqref{secrecycap}, the next step is to find a closed form expression and choose appropriate, optimal values for $T_{\rm s}, T_{\rm k}, Q$ to maximize this expression. This is shown in the next few sections.

\section{Secret Transmission Optimization}
This section studies the optimization of the secret transmission throughput of the system in terms of finding the best parameters for $T_{\rm s}, T_{\rm k}, Q$. 

We formulate the optimization problem below as
\begin{align}
\label{optimalP1}
    & (P1)\hspace{0.1in}
   \underset{Q,T_{\rm k},T_{\rm s}}{\text{max}}\ \   {R}_{\rm S}\\
    \label{constraintlam1}
     \text{s.t.} &\hspace{0.2in}\ T_{\rm k} \in \mathcal{T},\\
          & \hspace{0.2in}\ 2\leq T_{\rm s} \leq T_{\rm k}  ,\\
            & \hspace{0.2in} Q\in \{2^K|K=1,2 \dots\}
\end{align}
where {$\mathcal{T}= \{2,\dots, 	T\}$. }
The optimization variables are deeply coupled in this problem which makes it hard to tackle. Thus, we decouple $(P1)$ where we optimize one parameter at a time in a {one-shot manner}.
\subsection{Optimizing Key Generation Time Allocation}

For the time dedicated for key generation $T_{\rm k}$, observe that it should be optimized such that 
the number of key bits and and data bits 
are closest to each other because of the OTP encryption method. Thus, we propose to choose $T_{\rm k}$ as 
\begin{align}
\label{tkprob1}
    T_{\rm k}^*=\underset{T_{\rm k} \in \mathcal{T}}{\text{argmin}}\hspace{0.1in}  \left|\frac{T_{\rm k}}{2T}\bar{R}_{\rm k}- \frac{ (T-{T_{\rm k}})}{T}R_{\rm I}^{(L)}\right|.
\end{align}  Note that $T_{\rm k}$ is an even integer. 
Using \eqref{eqravgscale}, we can rewrite \eqref{tkprob1}   as
\begin{align}
\label{tkprob12} 
    T_{\rm k}^*=\underset{T_{\rm k} \in \mathcal{T}}{\text{argmin}}\hspace{0.1in} \left |\frac{T_{\rm k}}{2T}\bar{R}_{\rm k}- \frac{ (T-{T_{\rm k}})}{T}(\tilde{R}_I+\log\log L)\right|,
\end{align}
in which 
\begin{align}
\tilde{R}_I=\log\bigg(\frac{(\beta_{\rm ab} + \beta_{ar}\beta_{rb}\|\mathbf{R}\|_F^2)P}{\sigma^2}\bigg).
\end{align}
Now, we can solve this numerically, where the optimal solution for \eqref{tkprob12}   is found by solving
\begin{align}
\label{numTk}
    (\bar{R}_{\rm k}+\tilde{R}_I){T}_{\rm k}+ \log\log(\frac{T_{\rm k}}{T_{\rm s}})({T}_{\rm k}-{T})=T \tilde{R}_I.
\end{align}

On the other hand, 
we can find a sub-optimal but closed-form solution for an approximation for \eqref{tkprob12} by noting that $\frac{T_{\rm k}}{2T}\bar{R}_{\rm k}$ is increasing linearly in $T_{\rm k}$ whilst $\frac{ (T-{T_{\rm k}})}{T}R_{\rm I}^{(L)}$ is decreasing {non-linearly} in $T_{\rm k}$ {since it also appears in the logarithm $R_{\rm I}^{(L)}$} . We can approximate the term  $\frac{ (T-{T_{\rm k}})}{T}R_{\rm I}^{(L)}$ as linear when $\frac{ (T-{T_{\rm k}})}{T}\log\log(L)$ is neglected in the expression {compared to the }linear component $\frac{ (T-{T_{\rm k}})}{T}\tilde{R}_I$. Thus, the optimization problem reduces to a simpler one of finding the intersection point between the lines $\frac{T_{\rm k}}{2T}\bar{R}_{\rm k}$ and $\frac{ (T-{T_{\rm k}})}{T}\tilde{R}_I$ leading to
\begin{align}
\label{optimalTk}
    \tilde{T}_{\rm k}^*= \frac{{T}\tilde{R}_I}{\bar{R}_{\rm k}+\tilde{R}_I}.
\end{align}

\subsection{Optimizing RIS Switching Interval}
Next, we focus on the switching interval for the RIS $T_{\rm s}$, increasing it {increases the match between Alice and Bob's key bits due to better channel estimation, but can also decrease the secret key rate desired to be large}. Both quantities are encapsulated in the objective function. 
\begin{lemma}
\label{lemmats}
The optimal solution for $T_{\rm s}^*$ is 
\begin{align}
   T_{\rm s}^*= \underset{ 2\leq T_{\rm s} \leq T_{\rm k} }{\text{argmax}}{\hspace{0.1in}{R}_{\rm S}}=2
\end{align}
\end{lemma}
\begin{IEEEproof}
  First, $\bar{R}_{\rm k}$ increases when $T_s$ decreases and $R_{\rm I}^{(L)}$ depends on $T_{\rm s}$ inversely via $\log\log\frac{1}{T_{\rm s}}$. The minimum decreasing functions is decreasing .  
  Finally, since both are positive, decreasing functions, the product of them is also decreasing in $T_s$.
Hence, the optimal value for $T_s=2$.
\end{IEEEproof}

\subsection{Optimizing Quantization Levels}
Plugging $ T_{\rm k}^*$ and the optimal solution  $T_s=2$ back into \eqref{optimalP1} presents a simpler problem {where $R_{\rm S}=\frac{T_{\rm k}^*}{2T}\bar{R}_{\rm k}$ (or $R_{\rm S}=\frac{ (T-{T_{\rm k}^*})}{T}R_{\rm I}^{(L)}$) because $T_{\rm k}^*$ }guarantees equality between the two terms.
Thus, only $Q$ remains to be optimized. Observe that only $\frac{T_{\rm k}^*}{2T}\bar{R}_{\rm k}$ depends on $Q$, so we choose $Q^*$ using exhaustive search based on
\begin{align}
    Q^*= \underset{ Q\in \{2^K|K=1,2 \dots\}}{\text{argmax}}\frac{\bar{R}_{\rm k}R_{\rm I}^{(L)}}{2(\bar{R}_{\rm k}+R_{\rm I}^{(L)})},
\end{align}
for $\bar{R}_{\rm k}$ as given in \eqref{avgthrough} and $R_{\rm I}^{(L)}$ as given in \eqref{eqravgscale}.

Next, we provide numerical evaluations that illustrate the effect of the RIS on the key generation in terms of several characteristics: switching rate, size, inter-element correlation, location, and phases.

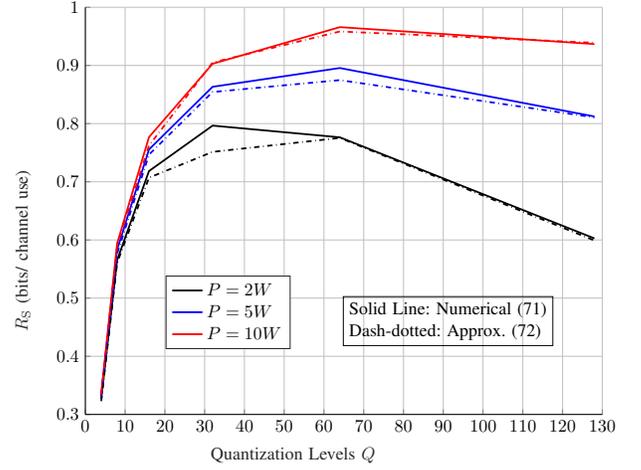
\begin{figure}
\tikzset{every picture/.style={scale=0.67}, every node/.style={scale=1}}
%
%
\begin{tikzpicture}

\begin{axis}[%
width=6.028in,
height=4.751in,
at={(1.011in,0.644in)},
scale only axis,
xmin=0,
xmax=130,
xlabel style={at={(0.6,0)}, font=\color{white!15!black}},
xlabel={Quantization Levels $Q$},
ymin=0.3,
ymax=1,
ylabel style={at={(0,0.6)},font=\color{white!15!black}},
ylabel={${R}_{\rm S}$  (bits/ channel use)},
axis background/.style={fill=white},
axis x line*=bottom,
axis y line*=left,
xmajorgrids,
ymajorgrids,
legend style={at={(0.232,0.24)}, anchor=south west, legend cell align=left, align=left, draw=white!15!black}
]
\addplot [color=black, line width=1.0pt]
  table[row sep=crcr]{%
4	0.326898630939572\\
8	0.56534236436258\\
16	0.718395599762691\\
32	0.796564491706553\\
64	0.776437547197466\\
128	0.602696545297942\\
};
\addlegendentry{$P=2 W$}

\addplot [color=blue, line width=1.0pt]
  table[row sep=crcr]{%
4	0.333242969552221\\
8	0.58364430440019\\
16	0.755520657273822\\
32	0.863326040737625\\
64	0.895564835762436\\
128	0.8122801983664\\
};
\addlegendentry{$P=5 W$}

\addplot [color=red, line width=1.0pt]
  table[row sep=crcr]{%
4	0.337017992384838\\
8	0.5942388614081\\
16	0.776703290218534\\
32	0.90289269117377\\
64	0.965882949800079\\
128	0.93677276588616\\
};
\addlegendentry{$P=10 W$}

\addplot [color=black, dashdotted, line width=1.0pt, forget plot]
  table[row sep=crcr]{%
4	0.322853566444196\\
8	0.560756733321875\\
16	0.707003721228176\\
32	0.75143203214064\\
64	0.775406326110041\\
128	0.599138900251598\\
};
\addplot [color=blue, dashdotted, line width=1.0pt, forget plot]
  table[row sep=crcr]{%
4	0.331357324650262\\
8	0.576452081424328\\
16	0.7466326097241\\
32	0.854354623086359\\
64	0.874801904662271\\
128	0.810667202007598\\
};
\addplot [color=red, dashdotted, line width=1.0pt, forget plot]
  table[row sep=crcr]{%
4	0.334678188077248\\
8	0.590075964771081\\
16	0.762174899633835\\
32	0.905052311194244\\
64	0.958412755657867\\
128	0.939005746175538\\
};
\end{axis}

\begin{axis}[%
width=7.778in,
height=5.833in,
at={(0in,0in)},
scale only axis,
xmin=0,
xmax=1,
ymin=0,
ymax=1,
axis line style={draw=none},
ticks=none,
axis x line*=bottom,
axis y line*=left
]
\node[below right, align=left, draw=black]
at (rel axis cs:0.5802,0.401) {Solid Line: Numerical \eqref{numTk}\\Dash-dotted: Approx. \eqref{optimalTk}};
\end{axis}
\end{tikzpicture}%
\caption{{Numerical solution for optimal $T_{\rm k}$ in comparison to the approximation in \eqref{optimalTk} for secret rate transmission versus Q}. {Simulation  parameters are found in Table \ref{tabsims}.}}
    \label{fig:trialTk}
\end{figure}

\section{Numerical Evaluation}

To evaluate the performance of the proposed protocol, we simulate it for a system with parameters provided in Table \ref{tabsims}. 
     Furthermore, we highlight the performance of our proposed scheme against a scheme without the RIS.
 



{Fig. \ref{fig:trialTk} shows the secrecy rate versus $Q$ for a number of transmit power levels. We observe how the optimal quantization resolution increases as the power or SNR  increases. This is due in part to a smaller estimation error in the channel estimates. Furthermore, the approximate solution for $T_{\rm k}$ in \eqref{optimalTk} yields a similar performance to that in \eqref{numTk}} 


   \begin{figure}[t]
\tikzset{every picture/.style={scale=0.68}, every node/.style={scale=1}}
%
%
\begin{tikzpicture}

\begin{axis}[%
width=6.028in,
height=4.751in,
at={(1.011in,0.644in)},
scale only axis,
xmin=2,
xmax=20,
xlabel style={at={(0.6,0)}, font=\color{white!15!black}},
xlabel={$T_{\rm s}$},
ymin=0,
ymax=7,
ylabel style={at={(0,0.6)}, font=\color{white!15!black}},
ylabel={$R_{\rm S}$  (bits/ channel use)},
axis background/.style={fill=white},
axis x line*=bottom,
axis y line*=left,
xmajorgrids,
ymajorgrids,
    legend style={at={(0.747,1.204)}, anchor=north west, legend cell align=left, align=left, draw=white!15!black}
]
\addplot [color=black, line width=1.0pt, mark=o, mark options={solid, black}]
  table[row sep=crcr]{%
2	6.90532323881256\\
4	4.71225991408923\\
6	3.61567656048198\\
8	2.94939653121188\\
10	2.49884835314378\\
12	2.17260635579957\\
14	1.92482582432309\\
16	1.72988043175861\\
18	1.57228044428572\\
20	1.44209202096132\\
};
\addlegendentry{SKR \eqref{SKR_lb}, $T_{\rm k}^* $.}

\addplot [color=black, dashed, line width=1.0pt]
  table[row sep=crcr]{%
2	6.29003774293222\\
4	3.3942895568346\\
6	2.36034852994526\\
8	1.8221392380724\\
10	1.48990331347872\\
12	1.26350513590093\\
14	1.09888927363166\\
16	0.973568233925067\\
18	0.874834126767066\\
20	0.794950770265046\\
};
\addlegendentry{SKR \eqref{SKR_lb}, $T_{\rm k}=\frac{T}{2} $.}

\addplot [color=blue, line width=1.0pt, mark=o, mark options={solid, blue}]
  table[row sep=crcr]{%
2	2.72519648370065\\
4	1.61626317675966\\
6	1.18230813903716\\
8	0.93891562515384\\
10	0.785941518577857\\
12	0.674925158964835\\
14	0.595615642723434\\
16	0.520965834517321\\
18	0.497643009173598\\
20	0.435526051370547\\
};
\addlegendentry{ESKR \eqref{avgthrough},  $T_{\rm k}^*, Q^*$}

\addplot [color=blue, dashed, line width=1.0pt]
  table[row sep=crcr]{%
2	1.67299912211226\\
4	0.898157402152635\\
6	0.615057741519106\\
8	0.468029462927146\\
10	0.377893747396219\\
12	0.316950955683342\\
14	0.272981372444063\\
16	0.239753813366073\\
18	0.213756317457378\\
20	0.192858248216695\\
};
\addlegendentry{ESKR \eqref{avgthrough},  $T_{\rm k}^*, Q=4$}

\addplot [color=red, line width=1.0pt]
  table[row sep=crcr]{%
2	0.0611157046769528\\
4	0.063994319789391\\
6	0.0637111821863789\\
8	0.0643303734244465\\
10	0.0647227243252774\\
12	0.0655385502528305\\
14	0.0653302445715508\\
16	0.0649728908005186\\
18	0.0657501476608382\\
20	0.0661844732564616\\
};
\addlegendentry{No RIS}

\end{axis}

\begin{axis}[%
width=7.778in,
height=5.833in,
at={(0in,0in)},
scale only axis,
xmin=0,
xmax=1,
ymin=0,
ymax=1,
axis line style={draw=none},
ticks=none,
axis x line*=bottom,
axis y line*=left
]
\end{axis}
\end{tikzpicture}%
  \caption{$R_{\rm s}$ versus $T_{\rm s}$ for validation of results and observing the effects of optimization.}
    \label{fig:rsvsts}
\end{figure}
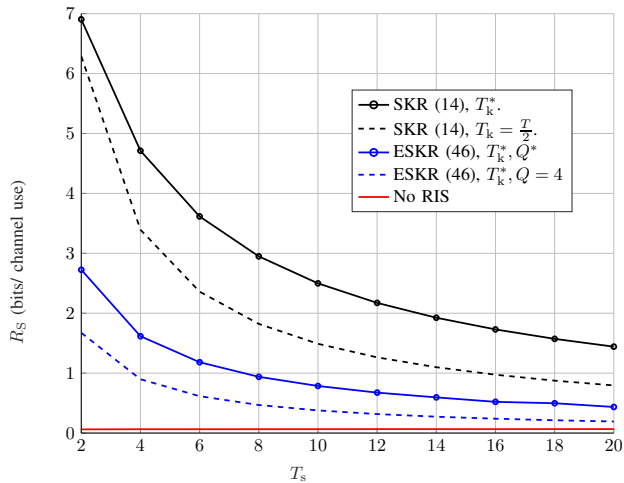

Fig. \ref{fig:rsvsts} shows the secrecy rate $R_S$ versus the number of symbols for RIS switching interval $T_{\rm s}$. First, we simulate the secrecy rate under theoretical SKR found in \eqref{rkmlb} with the RIS configuration chosen based on opportunistic beamforming as in \eqref{eqRISmax}. We compare with $R_S$ under the ESKR \eqref{avgthrough} which arises when we assume a non-zero mismatch probability for the keys and provide a practical key extraction protocol and introduce quantization levels $Q$. { The effect of choosing the optimal value for parameter $T_{\rm k}$ and $Q$ is apparent, as there is an increasing gap.  Overall, the secrecy rate decreases as $T_{\rm s}$ increases, implying that a higher RIS switching rate improves the performance of the rate which validates Lemma \ref{lemmats}.}
Moreover, one can see that if we choose fixed, non-optimal values for $Q$ and  $T_{\rm k}$, the secrecy rate degrades notably. {Observe how as $T_{\rm s}$ increases, the performance of the scheme with the RIS resembles that without the RIS. The effect of the RIS decreases marginally if it does not change its configuration within the key generating allocated time. }

 \begin{figure}[t]
\tikzset{every picture/.style={scale=0.65}, every node/.style={scale=1}}
%
%
\begin{tikzpicture}

\begin{axis}[%
width=6.028in,
height=4.751in,
at={(1.011in,0.644in)},
scale only axis,
xmin=5,
xmax=50,
xlabel style={at={(0.6,0)}, font=\color{white!15!black}},
xlabel={Eve distance from Alice $d_{AE}$},
ymin=0,
ymax=10,
ylabel style={at={(0,0.6)},font=\color{white!15!black}},
ylabel={$R_{\rm S}$ (bits/ channel use)},
axis background/.style={fill=white},
axis x line*=bottom,
axis y line*=left,
xmajorgrids,
ymajorgrids,
legend style={at={(0.171,0.672)}, anchor=south west, legend cell align=left, align=left, draw=white!15!black}
]
\addplot [color=black, line width=1.0pt]
  table[row sep=crcr]{%
2	6.82891179229284\\
5	6.82890393506555\\
8	6.82889611381049\\
11	6.82888759567762\\
14	6.82887724055358\\
17	6.82886323809189\\
20	6.82884184457877\\
23	6.82880306082766\\
26	6.8287067814605\\
29	6.82801070226473\\
32	6.82276522728908\\
35	6.82648323365376\\
38	6.82740858314583\\
41	6.82782953797259\\
44	6.82807048619085\\
47	6.82822664886383\\
50	6.82833610157965\\
};
\addlegendentry{$R_{\rm S}$ with SKR \eqref{SKR_lb}}

\addplot [color=blue, line width=1.0pt]
  table[row sep=crcr]{%
2	2.65264165789843\\
5	2.7204310295377\\
8	2.6733478967366\\
11	2.66427260743472\\
14	2.71648071462758\\
17	2.69157560450574\\
20	2.69941934584433\\
23	2.683751052505\\
26	2.6243841498467\\
29	2.66556751722175\\
32	2.60652807061556\\
35	2.72306732034791\\
38	2.66297821586027\\
41	2.73363460215489\\
44	2.67464645322548\\
47	2.69811071629177\\
50	2.65264165789843\\
};
\addlegendentry{$R_{\rm S}$ with ESKR \eqref{avgthrough}}

\addplot [color=red, line width=1.0pt]
  table[row sep=crcr]{%
2	0.092384036247608\\
5	0.0916764788642269\\
8	0.0897941619238455\\
11	0.0886275530959648\\
14	0.0919480707282177\\
17	0.0921113527091587\\
20	0.0915680322187396\\
23	0.0903822273163828\\
26	0.0904894927362429\\
29	0.0914055653228924\\
32	0.0921658346605646\\
35	0.0930962332630136\\
38	0.0922748806482028\\
41	0.0936472691122324\\
44	0.0907581223788645\\
47	0.0896343342963037\\
50	0.0906505905037735\\
};
\addlegendentry{$R_{\rm S}^{D1}$ no RIS (Key) \eqref{secrecyd1}}

\addplot [color=black, line width=1.0pt, mark=o, mark options={solid, black}]
  table[row sep=crcr]{%
2	0\\
5	0\\
8	0\\
11	0\\
14	0\\
17	0\\
20	0\\
23	0\\
26	0\\
29	0\\
32	0.360849347867129\\
35	0.835496316943319\\
38	1.31666523264066\\
41	1.58592498367364\\
44	2.11907515182057\\
47	2.37439676649048\\
50	2.66921615238807\\
};
\addlegendentry{$R_{\rm S}^{D2}$ no RIS  (Keyless) \eqref{secrecyd2}}

\addplot [color=black, dashed, line width=1.0pt,forget plot]
  table[row sep=crcr]{%
2	7.49058362187713\\
5	7.49057687037944\\
8	7.49057063128753\\
11	7.49056390765455\\
14	7.49055561208115\\
17	7.49054413538803\\
20	7.49052621517314\\
23	7.49049317990906\\
26	7.4904102787704\\
29	7.4898077067614\\
32	7.48537098902559\\
35	7.48858044884419\\
38	7.48937624704316\\
41	7.48973609489405\\
44	7.48994056214736\\
47	7.49007200179991\\
50	7.49016332756456\\
};
\addplot [color=blue,, dashed, line width=1.0pt, forget plot]
  table[row sep=crcr]{%
2	3.03451420573085\\
5	3.04795672471949\\
8	3.0978450131679\\
11	3.0360044715602\\
14	3.01373684015354\\
17	3.07353673268687\\
20	3.07656272122324\\
23	3.03153616363435\\
26	3.09174625514915\\
29	3.02262181937995\\
32	3.0360044715602\\
35	3.03004838360908\\
38	3.03004838360908\\
41	3.02558998527063\\
44	3.05996324464663\\
47	3.0978450131679\\
50	3.04795672471949\\
};
\addplot [color=red, dashed, line width=1.0pt,  forget plot]
  table[row sep=crcr]{%
2	0.126468173598736\\
5	0.125796275714636\\
8	0.125796275714636\\
11	0.125647557574731\\
14	0.1253507613259\\
17	0.126468173598736\\
20	0.124465433698842\\
23	0.124612465815898\\
26	0.126318483664573\\
29	0.12557327866502\\
32	0.126618081862986\\
35	0.123660438978233\\
38	0.125128720902925\\
41	0.12729539465741\\
44	0.128281739730155\\
47	0.127069125473073\\
50	0.126169011188053\\
};
\addplot [color=black, dashed, line width=1.0pt, mark=o, mark options={solid, black}, forget plot]
  table[row sep=crcr]{%
2	0\\
5	0\\
8	0\\
11	0\\
14	0\\
17	0\\
20	0.0453729321499701\\
23	0.740220340767447\\
26	1.41174642878517\\
29	1.9450270540836\\
32	2.56399582185438\\
35	2.97911742886268\\
38	3.24501111551609\\
41	3.85685994574961\\
44	4.10706891802139\\
47	4.5411410337634\\
50	4.78469778460767\\
};
\addplot [color=black, dotted, line width=1.0pt,  forget plot]
  table[row sep=crcr]{%
2	9.1649835022413\\
5	9.16498010840836\\
8	9.16497720377635\\
11	9.16497390095018\\
14	9.16496947965462\\
17	9.1649629040269\\
20	9.16495207721113\\
23	9.16493141631313\\
26	9.16487852223304\\
29	9.16449055743118\\
32	9.16172085730548\\
35	9.16377707250668\\
38	9.16428433521931\\
41	9.16451168976016\\
44	9.16463948471401\\
47	9.1647206530357\\
50	9.16477633100609\\
};
\addplot [color=blue, dotted, line width=1.0pt,  forget plot]
  table[row sep=crcr]{%
2	3.92649015448845\\
5	3.92451582446291\\
8	3.91794249335473\\
11	3.95829349546765\\
14	3.95829349546765\\
17	3.93836877503689\\
20	3.94600633727157\\
23	3.91663383946004\\
26	3.980400548785\\
29	3.97233808994753\\
32	3.92254303353442\\
35	3.96029489046518\\
38	3.90877612585563\\
41	3.98444203998857\\
44	3.99254582841418\\
47	3.97426340753068\\
50	3.96229793311285\\
};
\addplot [color=red, dotted, line width=1.0pt,  forget plot]
  table[row sep=crcr]{%
2	0.160198187356621\\
5	0.159652944569808\\
8	0.157938106941425\\
11	0.158789989314432\\
14	0.160088765706351\\
17	0.15889724109673\\
20	0.163114033646825\\
23	0.160307798253903\\
26	0.161414631201347\\
29	0.159870484706633\\
32	0.158789989314432\\
35	0.159544449069299\\
38	0.159004666964941\\
41	0.158044006290281\\
44	0.16108049248179\\
47	0.157938106941425\\
50	0.159761622686156\\
};
\addplot [color=black, dotted, line width=1.0pt, mark=o, mark options={solid, black}, forget plot]
  table[row sep=crcr]{%
2	0\\
5	0\\
8	0\\
11	0.388730949733972\\
14	1.76689955748484\\
17	2.78679687065597\\
20	3.75178695461981\\
23	4.39851991568399\\
26	5.04429099113464\\
29	5.47418757910713\\
32	6.1780971112475\\
35	6.65375055465336\\
38	6.93412304187745\\
41	7.45162284698274\\
44	7.8226633843766\\
47	8.05415981746976\\
50	8.52475650354856\\
};
\end{axis}

\begin{axis}[%
width=7.778in,
height=5.833in,
at={(0in,0in)},
scale only axis,
xmin=0,
xmax=1,
ymin=0,
ymax=1,
axis line style={draw=none},
ticks=none,
axis x line*=bottom,
axis y line*=left
]
\node[below right, align=left, draw=black]
at (rel axis cs:0.2702,0.9051) {Dotted:  $d_{AB}=10$\\ Dashed: $d_{AB}=20$ \\ Solid: $d_{AB}=30$ };
\end{axis}
\end{tikzpicture}%
    \caption{{Secrecy Rate for RIS vesus direct channel only against Eve's horizontal distance from Alice for different Bob positions. Here, $d_{AB}$ is the horizontal distance between Alice and Bob. }}
    \label{fig:direct}
\end{figure}
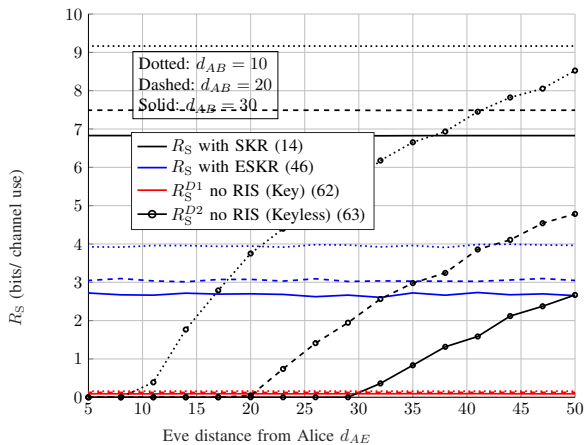
Next, we investigate the role of the RIS in enhancing secret transmission by comparing with all possible secrecy transmission scenarios without an RIS, i.e. keyless using a wiretap channel model \cite{wiretapWyner} and with key using our proposed protocol without an RIS. {We also consider different distances for Bob and Eve from Alice in order to investigate the effect of distance on the overall secrecy rate. }One can observe that keyless security is reliant heavily on the positions of all the users. As Eve moves away from Alice, the secrecy rate increases due to a decrease in secrecy leakage. Moreover, as Bob moves closer to Alice their channel gain increases which also increases the secrecy rate. On the other hand, the key-based secrecy rate is impertinent to Eve's position as long as Eve is spatially decorrelated from both users. When Bob moves closer to Alice,  a noticeable increase is incurred in the key-based secrecy rate due to the channel gain's effect which is encapsulated in the match probability in key rate and in the ergodic  information rate.  { We can conclude that secrecy rate under the RIS performs better in certain scenarios when Eve is in proximity with Alice or Bob as key-less security would be pointless. {Thus, the use of the RIS in our key-based scheme must be the go-to protocol when the RIS is located close to Alice or Bob and when Alice to Eve's channel is stronger than Alice to Bob's channel.} }

\section{Conclusion}
 In this paper, we study the RIS effect in enhancing key generation and secure transmission, where the RIS provides a two-fold enhancement by adding additional channels and by perturbing the static channel in order to obtain a higher key rate. We formulate an expression for the theoretical achievable SKR lower bound using our proposed system model under block fading channels. Moreover, we derive the average key throughput for a proposed protocol and the secure information transmission.  Furthermore, we study the effect of changing RIS parameters such as the number of elements and the switching rate of the RIS, as well as system parameters such as the quantization resolution $Q$ and key generation time allocation  on the overall secrecy rate of the system. { The results show that the RIS can improve the secrecy rate when it is adjacent to Alice or Bob, and when the Alice to Bob's channel is stronger than Alice to Eve's channel. The advantages of the scheme is that secrecy with RIS is virtually guaranteed despite the relative positions of Bob and Eve to Alice, and that the secrecy rate can grow not only by relying on fixed channel characteristics and SNR, but also with a new degree-of-freedom which is the switching rate of the RIS.}


\bibliographystyle{IEEEtran}
\bibliography{bib.bib}

\end{document}